\documentclass[conference]{IEEEtran}

\IEEEoverridecommandlockouts

\usepackage{url}
\usepackage{array}
\usepackage{cancel}
\usepackage{siunitx}
\usepackage{multirow}
\usepackage{graphicx}
\usepackage{textcomp}
\usepackage{multirow}
\usepackage{stfloats}
\usepackage{listings}
\usepackage{colortbl}
\usepackage{algorithm}
\usepackage[T1]{fontenc}
\usepackage[10pt]{moresize}
\usepackage[utf8]{inputenc}
\usepackage[noadjust]{cite}
\usepackage{mathtools, cuted}
\usepackage[noend]{algpseudocode}
\usepackage{amsmath,bm,amssymb,amsfonts}
\usepackage{csvsimple}
\usepackage{soul}
\usepackage{arydshln}

\usepackage{tikz}
\usepackage{tikzsymbols}
\usetikzlibrary{spy, positioning, shapes, arrows, chains, patterns,decorations.markings,pgfplots.groupplots}

\def\BibTeX{{\rm B\kern-.05em{\sc i\kern-.025em b}\kern-.08em
		T\kern-.1667em\lower.7ex\hbox{E}\kern-.125emX}
}

\graphicspath{ {figs} }

\begin{document}
\author{
  \IEEEauthorblockN{Carlos Gewehr\textsuperscript{*}}
  \IEEEauthorblockA{Electrical and Computer Engineering\\
    Carnegie Mellon University\\
    Pittsburgh, PA\\
    cdearauj@andrew.cmu.edu}
  \and
  \IEEEauthorblockN{Levent Aksoy\textsuperscript{*}}
  \IEEEauthorblockA{Computer Systems\\
    Tallinn University of Technology\\
    Tallinn, Estonia \\
    levent.aksoy@taltech.ee}
  \and
  \IEEEauthorblockN{Mariia Rodinko}
  \IEEEauthorblockA{Intelligent Software Systems and Technologies\\
    V. N. Karazin Kharkiv National University\\
    Kharkiv, Ukraine \\
    mariia.rodinko@karazin.ua}
  \and
  \IEEEauthorblockN{ \hspace{40pt}Roman Oliynykov}
  \IEEEauthorblockA{ \hspace{40pt}Cybersecurity of Inf. Systems, Networks and Technologies\\
     \hspace{40pt}V. N. Karazin Kharkiv National University\\
     \hspace{40pt}Kharkiv, Ukraine \\
     \hspace{40pt}roman.oliynykov@karazin.ua}
  \and
  \IEEEauthorblockN{Samuel Pagliarini}
  \IEEEauthorblockA{Electrical and Computer Engineering\\
    Carnegie Mellon University \\
    Pittsburgh, PA \\
    pagliarini@cmu.edu}
}

	\title{Kalyna Block Cipher: From Design Space Exploration to ASIC Design}
	
	\maketitle

\begingroup
    \renewcommand\thefootnote{*}
    \footnotetext{These authors contributed equally to this work.}
\endgroup
	
	\begin{abstract}
		The Kalyna block cipher is a Ukrainian cryptography standard, selected through a national competition held between 2007 and 2010 and approved in 2015. Although its software implementations have been introduced, hardware-efficient implementations of the algorithm, i.e., accelerators, do not exist. In this paper, we explore various design architectures to implement its encryption, decryption, and unified encryption/decryption functions, considering the trade-off between area and latency. We present hardware reduction techniques and introduce alternative designs with low area, latency, and energy consumption, targeting an application-specific integrated circuit (ASIC). We present hardware-efficient designs that include countermeasures against side-channel analysis (SCA) and fault injection (FI) attacks, such as hiding, masking, and duplication techniques. We validate these implementations in a 65\;nm  ASIC chip. Experimental results confirm the need for alternative designs that explore the design search space for different requirements. The proposed architectures enable hiding the power SCA leakage by randomizing the execution of operations, and the temporal duplication in designs with countermeasures against the SCA attacks can mitigate the FI attacks. The functionality of the ASIC test chip, including various Kalyna designs, is validated through measurements.
	\end{abstract}
	
	
	\section{Introduction}

Block ciphers are symmetric cryptography primitives widely used for secure communication. They can also be used to provide confidentiality under certain modes and for various other functions: hash functions, message authentication codes, and pseudo-random number generators. Over the years, many efficient block ciphers have been introduced, standardized by national and international organizations, and used in our daily life applications~\cite{aes, present_iso, sonmez25_noetal}.

The GOST 28147-89 block cipher~\cite{gost} was used as the main block cipher in Ukraine until 2015. 
Since its software implementation on scalar general-purpose CPU (without complex SIMD extensions like AVX) is slow when compared to other standards, such as AES~\cite{aes}, and effective theoretical attacks were discovered, it was replaced by the Kalyna block cipher~\cite{kalyna_noetal} selected through a national public cryptographic competition. Among other candidates, Kalyna offered a compact implementation and performance improvement. It is based on the substitution-permutation network (SPN) with an increased maximum distance separable (MDS) matrix size, four different S-boxes, pre- and post-whitening using $2^{64}$ modulo addition, and a prominent key schedule construction. 

Although there are software implementations of Kalyna~\cite{zaiats22} and the algorithm has been deployed in practice, efforts to create hardware accelerators for Kalyna are not known. Hence, in this work, we target the implementation of the Kalyna block cipher, considering all its possible functions, i.e., encryption, decryption, and unified encryption/decryption, with all valid block sizes and key lengths, targeting an application-specific integrated circuit (ASIC) as a design platform. We introduce parallel and serial design architectures, where the trade-off between area and latency is explored. We propose hardware reduction techniques in its fundamental operations, such as the byte substitution and mix column operations, offer a fast addition operation, such as the Kogge-Stone adder (KSA)~\cite{han87}, and present design architectures that use a small number of operations, sharing common resources. On the design architecture that has the least area requirement, we introduce countermeasures against the side-channel analysis (SCA) attacks~\cite{standaert10} using hiding and masking techniques. The same design is also protected against fault injection (FI) attacks~\cite{barenghi12} by duplication techniques. Finally, we validate various Kalyna designs in an ASIC chip under the 65\;nm technology. Experimental results show that the proposed architectures lead to alternative designs with low area, latency, and energy consumption, enabling a designer to choose the one that best fits the specific requirements of an application. The proposed hardware reduction techniques can reduce area, latency, and energy consumption significantly. The countermeasures used to mitigate the SCA and fault mitigation attacks can reduce the power SCA leakage confirmed by the test vector leakage assessment (TVLA)~\cite{goodwill11}, increasing the hardware complexity for the sake of security. The ASIC implementations validated in silicon show similar results when compared to the pre-silicon results.

The rest of this paper is organized as follows: Section~\ref{sec:background} describes the Kalyna block cipher.  Section~\ref{sec:architectures} presents the design architectures and the countermeasures against both SCA and FI attacks. Experimental results related to front-end and back-end designs are given in Section~\ref{sec:results}, results related to fabricated chips are presented in Section~\ref{sec:silicon}, and finally, the paper is concluded in Section~\ref{sec:conclusions}.

	\section{Background}
\label{sec:background}

This section describes the main functions of the Kalyna block cipher and its building blocks. 

\begin{figure}[t]
	\centerline{\includegraphics[width=8.0cm]{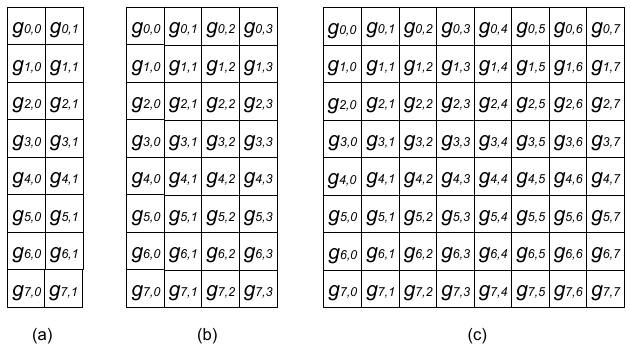}}
	\vspace*{-2mm}
	\caption{ISMs: (a)~when $l$ is 128; (b)~when $l$ is 256; (c)~when $l$ is 512.}
	\label{fig:ism}
	\vspace*{-6mm}
\end{figure}

The encryption transformation in the Kalyna block cipher is a mapping $E_{l,k}^{K} : V_{l} \rightarrow V_{l}$, where $K$ is the key, $l$ and $k$ are the block size and key length, respectively with $l, k \in \{128, 256, 512\}$, such that $k=l$ or $k=2l$, and $V_{j}$ is a \mbox{$j$-dimensional} vector space over the Galois field GF with $j \geq 1$. While its inputs are \mbox{$l$-bit} plaintext and $k$-bit key, its output is $l$-bit ciphertext. This transformation is implemented as iterative rounds of several operations, where the input of these operations, denoted as $x$, including plaintext and round key, is represented as an $8 \times c$ internal state matrix (ISM) where $x \in V_l$ and $c$ is the number of columns in ISM. Each entity of the ISM $G$ is denoted as $g_{i,j}$, where $g_{i,j} \in V_8$ with $0 \leq i \leq 7$ and $0 \leq j \leq c-1$. Table~\ref{tab:kalyna} presents the parameters and their values in the Kalyna block cipher, and Fig.~\ref{fig:ism} presents the ISMs under all possible block sizes.

\begin{table}[t]
	\centering
	\caption{Parameters of the Kalyna block cipher.}
	\vspace{-3mm}
	\footnotesize
	\begin{tabular}{|c|c|c|@{\hskip3pt}c@{\hskip3pt}|}
		\hline
		Block Size ($l$) & Key Length ($k$) & \#Rounds ($t$) & \#Columns in ISM ($c$)\\ 
		\hline \hline
		\multirow{2}{*}{128}  & 128 & 10 & \multirow{2}{*}{2}  \\
		& 256 & 14 &                     \\
		\hline
		\multirow{2}{*}{256}  & 256 & 14 & \multirow{2}{*}{4}  \\
		& 512 & 18 &                     \\
		\hline
		512                   & 512 & 18 & 8                   \\
		\hline
	\end{tabular}
	\label{tab:kalyna}
	\vspace{-6mm}
\end{table}

The encryption transformation $E_{l,k}^K$ is defined as follows:
\begin{equation}
	E_{l,k}^{K} = \eta_l^{K_t} \circ \psi_l \circ \tau_l \circ \pi_l \circ \prod_{v=1}^{t-1} (\kappa_l^{K_v} \circ \psi_l \circ \tau_l \circ \pi_l) \circ \eta_l^{K_0}
	\label{eqn:kalyna_enc}
\end{equation}
where $\eta_l^{K_v}$ is the modulo $2^{64}$ addition of ISM with the round key $K_v$, where $0 \leq v \leq t$, $\pi_l$ is the byte substitution operation using four different S-boxes, $\tau_l$ is the permutation of elements in ISM using right circular shift, $\psi_l$ is the mix column operation, which multiplies ISM by a constant matrix, and $\kappa_l^{K_v}$ is the modulo 2 addition of ISM with the round key $K_v$. Note that $\Gamma \circ \Lambda$ denotes the sequential application of transformations, where $\Lambda$ is applied first.

The Kalyna S-boxes are constructed as non-linear permutations with high algebraic complexity, low differential uniformity, and high nonlinearity to resist linear, differential, and algebraic cryptanalysis. The deployment of multiple S-boxes is a design paradigm shared with several classical and contemporary symmetric primitives given as follows:
\begin{itemize}
    \item The foundational Data Encryption Standard~\cite{pub1999data} utilizes eight distinct $6 \times 4$ S-boxes ($S_1 \dots S_8$) within its Feistel network.
    \item The original GOST 28147-89 block cipher~\cite{gost} uses eight secret $4 \times 4$ S-boxes; later, several public sets of S-boxes for this cipher were published in different countries.
    \item Twofish~\cite{schneier1998twofish} uses four key-dependent $8 \times 8$ S-boxes.
    \item Anubis~\cite{rijmen2000anubis} and Khazad~\cite{biryukov2003analysis} are involutional SPN ciphers constructed using multi-layer heterogeneous substitution units derived from smaller $4 \times 4$ S-boxes.
    \item ARIA~\cite{kwon2003new} is standardized in South Korea and employs two alternating substitution layer types, each utilizing four distinct $8 \times 8$ S-boxes ($S_1, S_2, S_1^{-1}, S_2^{-1}$) in a different order.
\end{itemize}

The flexibility to employ various S-boxes in Kalyna enables the implementation of an additional long-term secret key, preserving communication security even under session key compromise. Deploying multiple distinct public S-boxes in Kalyna increases resistance to practical cryptanalysis, requiring a small memory overhead in a software implementation.

The modulo $2^{64}$ addition operation processes through the 64-bit columns of ISM $G = (g_{i,j})$ and of the round key, where $0 \leq i \leq 7$ and $0 \leq j \leq c-1$. The byte substitution operation processes through each element of ISM $G$ as $\pi_{i\;mod\;4}(g_{i,j})$, where \mbox{$\pi_s : V_8 \rightarrow V_8$} with $s \in \{0,1,2,3\}$, using 4 different \mbox{S-boxes.} The permutation operation rotates each row of ISM $G$, $G_i$ with \mbox{$0 \leq i \leq 7$}, right by $\lfloor i \cdot l/512 \rfloor$. The mix column operation processes through each column of ISM $G$ as $(\upsilon \ggg i) \otimes G_j$, where $\upsilon~=~\{0x01, 0x01, 0x05, 0x01, 0x08, 0x06, 0x07, 0x04\}$ is a vector that forms the circulant matrix with the MDS property and $G_j$ is the $j^{th}$ column of ISM. Note that $\otimes$ denotes the scalar vector multiplication over GF($2^8$) using the irreducible polynomial $x^8+x^4+x^3+x^2+1$. The modulo 2 addition operation processes through each element of ISM $G$ and of the round key using {\sc xor} operations.

Kalyna is explicitly designed for modern 64-bit processor architectures. While the modular addition is typically the core non-linear operation in addition-rotation-XOR (ARX) primitives, such as SPECK~\cite{beaulieu13} and Threefish~\cite{ferguson2010skein}, its inclusion in Kalyna's pre- and post-whitening layers introduces extra non-linear key-mixing dependencies. This breaks the strictly affine key-injection paradigm of conventional SPNs like AES.

The decryption transformation $D_{l,k}^{K}$ is the inverse of $E_{l,k}^{K}$ and is defined as follows:
\begin{equation}
	\begin{split}
			D_{l,k}^{K} = (\eta_{l}^{K_0})^{-1} & \circ \prod_{v=t-1}^{1} (\pi_l^{-1} \circ \tau_l^{-1} \circ \psi_l^{-1} \circ (\kappa_{l}^{K_v})^{-1}) \\
			& \circ \pi_l^{-1} \circ \tau_l^{-1} \circ \psi_l^{-1} \circ (\eta_{l}^{K_t})^{-1}
		\end{split}
	\label{eqn:kalyna_dec}
\end{equation}
where $\phi^{-1}$ denotes the inverse of the $\phi$ operation. While its inputs are $l$-bit ciphertext and $k$-bit key, its output is $l$-bit plaintext. The inverse of the modulo $2^{64}$ addition is the modulo $2^{64}$ subtraction. The inverse of the byte substitution operation processes through each element of ISM $G$ as $\pi_{i\;mod\;4}^{-1}(g_{i,j})$, where \mbox{$\pi_s^{-1} : V_8 \rightarrow V_8$} with $s \in \{0,1,2,3\}$, using 4 different \mbox{S-boxes.} The inverse of the permutation operation rotates each row of ISM $G$, $G_i$ with \mbox{$0 \leq i \leq 7$}, left by $\lfloor i \cdot l/512 \rfloor$. The inverse of the mix column operation processes through each column of ISM $G$ as $(\upsilon^{-1} \lll i) \otimes G_j$, where $\upsilon^{-1}=\{0xAD, 0x95, 0x76, 0xA8, 0x2F, 0x49, 0xD7, 0xCA\}$ is a vector that forms the circulant matrix with the MDS property and $G_j$ is the $j^{th}$ column of ISM. In this case, the irreducible polynomial is the same as the one used in encryption. The inverse of the modulo 2 addition is the same as itself.

The key schedule, which is common in both encryption and decryption, consists of three main steps: (i)~intermediate key generation; (ii)~round key generation with even indexes; and (iii)~round key generation with odd indexes. The intermediate key $K_{\sigma}$ is generated as follows:
\begin{equation}
	\Theta^{K} = \psi_l \circ \tau_l \circ \pi_l \circ \eta_l^{K_{\alpha}} \circ \psi_l \circ \tau_l \circ \pi_l \circ \kappa_l^{K_{\omega}} \circ \psi_l \circ \tau_l \circ \pi_l \circ \eta_l^{K_{\alpha}}
	\label{eqn:kalyna_int}
\end{equation}
where $K_{\alpha} = K_{\omega} = K$ when $k=l$ and $K_{\alpha} || K_{\omega} = K$  when $k=2l$. The input to the $\Theta^{K}$ transformation is a constant given as $(l+k+64)/64$. Thus, the transformation to obtain the even round key when $i \in \{0, 2, \ldots t\}$ is defined as follows:
\begin{equation}
	\Xi^{K} = \eta_l^{\varphi_i^{K_{\sigma}}} \circ \psi_l \circ \tau_l \circ \pi_l \circ \kappa_l^{\varphi_i^{K_{\sigma}}} \circ \psi_l \circ \tau_l \circ \pi_l \circ \eta_l^{\varphi_i^{K_{\sigma}}}
	\label{eqn:kalyna_odd}
\end{equation}
where $\varphi_i^{K_{\sigma}}$ returns the modulo $2^{64}$ addition of the intermediate key $K_{\sigma}$ with a constant shifted by the round index, and the input to the $\Xi^{K}$ transformation is the key $K$ rotated to the right based on the round index. Thus, the round key with an odd index is determined as $K_i~=~K_{i-1} \lll (24+l/4)$ when $i \in \{1, 3, \ldots t-1\}$. 

The Kalyna one-way key schedule is considered "strong" because it's designed to prevent recovery of the main encryption key even if one or more round keys are compromised. This one-way property is achieved by using two keys within the key generation procedure: the primary encryption key and the intermediate key derived from it.

	\section{Design Methodology}
\label{sec:architectures}

This section introduces parallel and serial design architectures for Kalyna, aiming to explore the trade-off between area and latency, and presents hardware reduction techniques proposed for its building blocks. It also describes countermeasures against the SCA and FI attacks implemented in a design architecture that leads to the smallest area.

\subsection{Design Architectures}
\label{subsec:da}

This subsection initially describes the design architectures realizing the encryption function and then, summarizes the changes in these architectures realizing the decryption and unified encryption/decryption functions.

\begin{figure}[t]
	\centerline{\includegraphics[width=5.0cm]{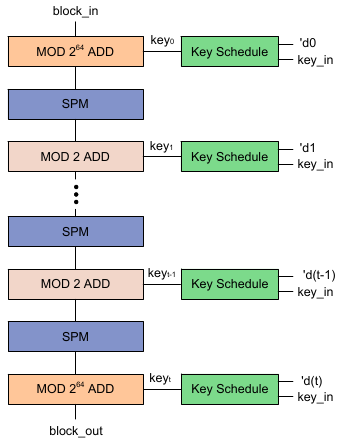}}
	\vspace*{-2mm}
	\caption{Parallel design architecture.}
	\label{fig:parallel}
	\vspace*{-4mm}
\end{figure}

First, we describe a parallel design architecture, denoted as \textit{parallel}, where all rounds are computed at the same time. The encryption process under this architecture is illustrated in Fig.~\ref{fig:parallel}, where \textit{block\_in}, \textit{key\_in}, and \textit{block\_out}, denote the plaintext, key, and ciphertext, respectively, and the SPM block includes the byte substitution, permutation, and mix column operations for the sake of simplicity, and the key schedule block performs the generation of the intermediate key and round keys with even and odd indexes. Note that the \textit{parallel} architecture leads to an extensive combinational circuit due to many necessary rounds. 

\begin{figure}[t]
	\centerline{\includegraphics[width=9cm]{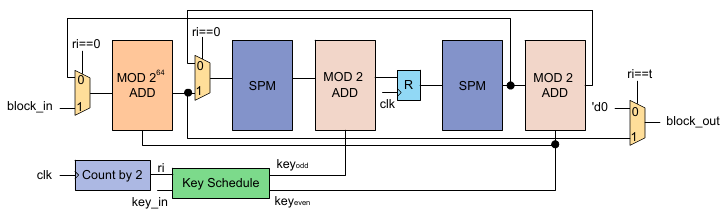}}
	\vspace*{-2mm}
	\caption{The \textit{serial2rnd} design architecture.}
	\label{fig:serial2rnd}
	\vspace*{-4mm}
\end{figure}

However, common operations in the \textit{parallel} architecture can be shared, reducing the hardware complexity at the expense of latency. The serial design architecture, where two rounds are realized in one clock cycle, denoted as \textit{serial2rnd}, is given in Fig.~\ref{fig:serial2rnd}. In this figure, $ri$ denotes the round index, and the counter counts from 0 to the number of rounds $t$ in steps of 2. Observe that the \textit{serial2rnd} design architecture uses a single key schedule block, which generates both round keys with even and odd indexes, two SPM and modulo 2 addition operations, and one modulo $2^{64}$ addition with a counter, register, and multiplexors to orchestrate the encryption process. Note that this architecture requires $t/2+1$ clock cycles to obtain the encrypted block, where the last clock cycle is used to store the encrypted block.

\begin{figure}[t]
	\centerline{\includegraphics[width=9cm]{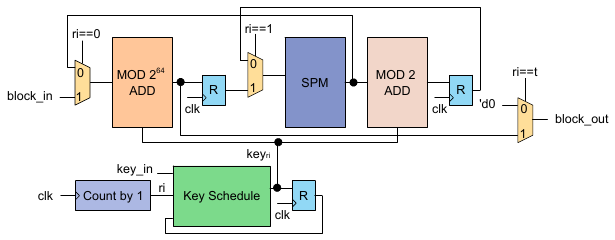}}
	\vspace*{-2mm}
	\caption{The \textit{serial1rnd} design architecture.}
	\label{fig:serial1rnd}
	\vspace*{-6mm}
\end{figure}

Moreover, each round of the encryption process can be implemented in one clock cycle using more control logic and registers, but only one modulo 2 and $2^{64}$ addition and SPM operation, as shown in Fig.~\ref{fig:serial1rnd}. This design architecture, denoted as \textit{serial1rnd}, uses one key schedule block and a counter that counts from 0 to the number of rounds $t$ in steps of 1. The round key with an even index needs to be stored to determine the round key with the next odd index, and the output of the modulo $2^{64}$ addition at round 0 needs to be stored for the input of the SPM block in the next round. This architecture requires $t+1$ clock cycles to obtain the encrypted block.

\begin{figure}[t]
	\centerline{\includegraphics[width=7.0cm]{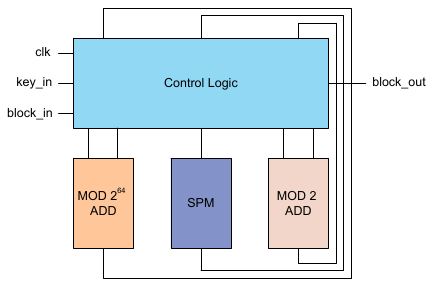}}
	\vspace*{-2mm}
	\caption{The \textit{serial\_1opr} design architecture.}
	\label{fig:serial1opr}
	\vspace*{-4mm}
\end{figure}

\begin{figure}[t]
	\centerline{\includegraphics[width=9.2cm]{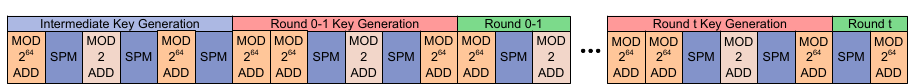}}
	\vspace*{-2mm}
	\caption{Schedule of operations in the \textit{serial1opr} design architecture.}
	\label{fig:schedule}
	\vspace*{-6mm}
\end{figure}

Furthermore, the hardware complexity of the encryption transformation can be reduced by sharing the common resources in the whole encryption process, including the key schedule and thus, only one modulo 2 and $2^{64}$ addition and SPM operation can be used in each clock cycle. Fig.~\ref{fig:serial1opr} illustrates this architecture denoted as \textit{serial1opr}, where the control logic includes a finite state machine to synchronize the process of each operation in each clock cycle, as shown in Fig.~\ref{fig:schedule}, and registers to store the round keys, round outputs, and the output of each operation. This architecture requires $(5t+13)+1$ clock cycles to obtain the encrypted block. To reduce the number of registers, these operations have the same input variable(s), since one operation is processed in each clock cycle.

\begin{figure}[t]
	\centerline{\includegraphics[width=9.2cm]{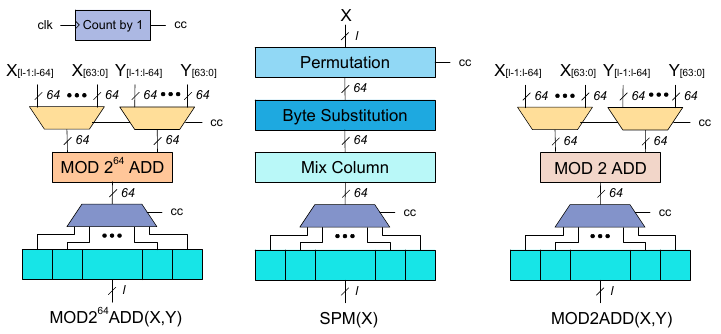}}
	\vspace*{-2mm}
	\caption{64-bit operations used in the \textit{serial1col} design architecture.}
	\label{fig:serial_1col}
	\vspace*{-4mm}
\end{figure}

Finally, rather than using the modulo 2 and $2^{64}$ addition and SPM operations handling $l$-bit computations in one clock cycle in the \textit{serial1opr} architecture, only the \mbox{64-bit} one column of ISM can be processed in each clock cycle. Fig.~\ref{fig:serial_1col} illustrates these operations processing 64 bits at a time. In this figure, the counter counts from 0 to the number of columns in ISM, i.e., $c$, in a step of 1. In the SPM block, the permutation operation is used before the byte substitution to handle the columns of ISM easily without changing the SPM functionality. This architecture is denoted as \textit{serial1col} and requires $(5t+13)(c+1)+1$ clock cycles. 

In the \textit{serial1opr} and \textit{serial1col} architectures, the modulo 2 and $2^{64}$ addition and SPM operations are controlled in such a way to execute only the required operation at the right time. 


The decryption transformation can be similarly implemented under the design architectures given for the encryption transformation. Additionally, inverses of the modulo $2^{64}$ addition and SPM operations are required. The unified Kalyna design, which can process encryption/decryption, can be implemented using common operations of both functions.

In order to reduce the hardware complexity further, a number of techniques are employed. For the modulo $2^{64}$ addition and its inverse, in addition to a generic adder (GA) described in Verilog, the fast and area-efficient KSA~\cite{han87}, selected among other adder architectures, was used. For the substitution operation and its inverse, the S-box designs with different numbers and sizes of multiplexors are described in Verilog, synthesized, and the one with the smallest hardware complexity was chosen. For the mix column operation and its inverse, similar to the work in~\cite{kranz17}, a tool that finds a small number of {\sc xor} gates to realize the constant matrix vector multiplication over GF($2^8$) was developed, and its solutions on those circulant matrices were used. 

\subsection{Security Countermeasures}
\label{subsec:si}

For the Kalyna block cipher with security countermeasures, the \textit{serial1col} design architecture is considered since it leads to designs with the smallest area as shown in Section~\ref{sec:results}. Our countermeasures focus on the well-known SCA and FI attacks. 

To mitigate the SCA attacks, two prominent techniques, called hiding and masking, are used. To hide the operation running according to the schedule given in Fig~\ref{fig:schedule}, in the encryption and decryption processes, two and four\footnote{While the encryption function includes 3 operations, i.e., modulo 2 and $2^{64}$ addition, and SPM, the decryption function uses 5 operations, i.e., the modulo 2 and $2^{64}$ addition, SPM, modulo $2^{64}$ subtraction, and inverse SPM.} other operations are randomly executed, respectively. In this case, the execution of unnecessary operation(s) is randomized based on a different single bit of ISM at each time. To randomize an operation, first-order masking is used in the modulo 2 addition, modulo $2^{64}$ addition/subtraction, and the byte substitution and mix column blocks of the SPM operation and its inverse. In this case, these blocks are initially implemented at gate-level using basic logic gates, e.g., {\sc and}, {\sc or}, {\sc xor}, {\sc not}, {\sc nand}, {\sc nor}, and {\sc xnor}, and then, Boolean masking~\cite{diehl16} is applied. The XorShift technique of~\cite{marsaglia03} is used as a pseudo-random number generator in the masked blocks. 

To mitigate FI attacks, operations are duplicated, and a control logic that checks if these two operations generate the same output is implemented. Since such a spatial duplication increases the design area significantly, a temporal duplication that duplicates the execution of operations in time is also implemented, doubling the number of clock cycles with a smaller area overhead than the spatial duplication. Moreover, a control logic that checks if the operations are executed in the given order is implemented to mitigate a fault injection attack that aims to alter the order of execution (i.e., control flow integrity). Note that whenever a faulty behavior is observed, the execution of the cipher is halted.

The Kalyna implementations are described in Verilog, considering all possible functions with all possible parameter values. They are freely available at~\cite{kalyna_github}.

	\section{Experimental Results}
\label{sec:results}

This section initially presents the \mbox{gate-level} synthesis results of the Kalyna block cipher designs under the architectures given in Section~\ref{subsec:da}, including all combinations of block size $l$ and key length $k$ implemented for the realization of the encryption, decryption, and unified encryption/decryption functions. Then, it presents the gate-level synthesis results of the designs, including countermeasures against SCA and FI attacks as discussed in Section~\ref{subsec:si}. Moreover, it introduces the physical design results of the Kalyna block cipher without and with security countermeasures. Finally, it introduces the pre-silicon test vector leakage assessment on selected designs.

\subsection{Designs without Security Countermeasures}

Table~\ref{tab:clock} presents the number of clock cycles required for the encryption and decryption processes under the given design architectures. Observe that while the \textit{serial1col} design architecture requires the highest number of clock cycles since it processes each column of ISM in each clock cycle, depending on both the number of rounds $t$ and the number of columns in ISM $c$, the number of clock cycles depends only on the number of rounds $t$ in other serial architectures.

\begin{figure*}[t]
	\centering
	\vspace{-6mm}
	\parbox{5.8cm}{\centerline{\includegraphics[width=6.5cm]{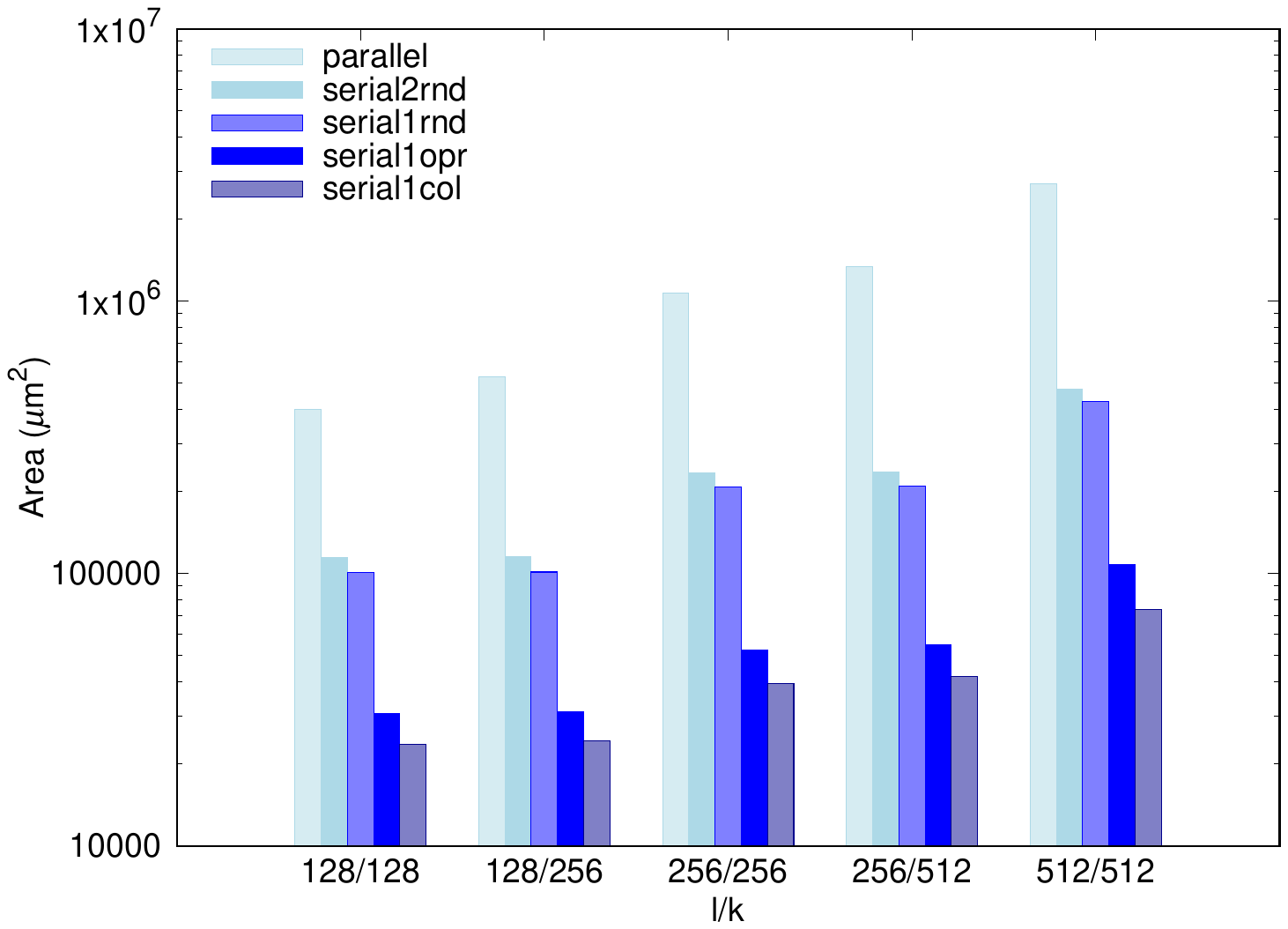}}}\
	\parbox{5.8cm}{\centerline{\includegraphics[width=6.5cm]{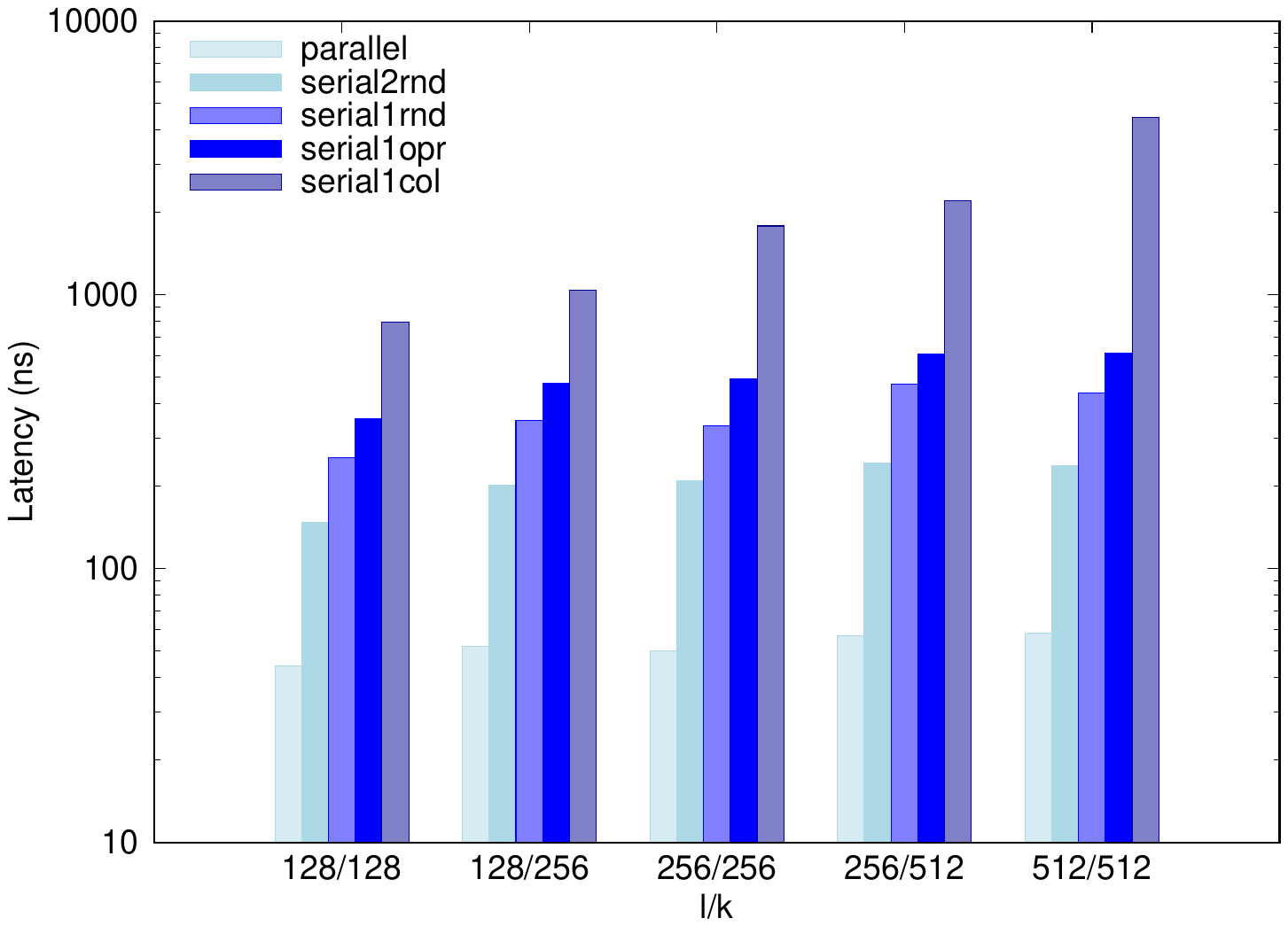}}}\
	\parbox{5.8cm}{\centerline{\includegraphics[width=6.5cm]{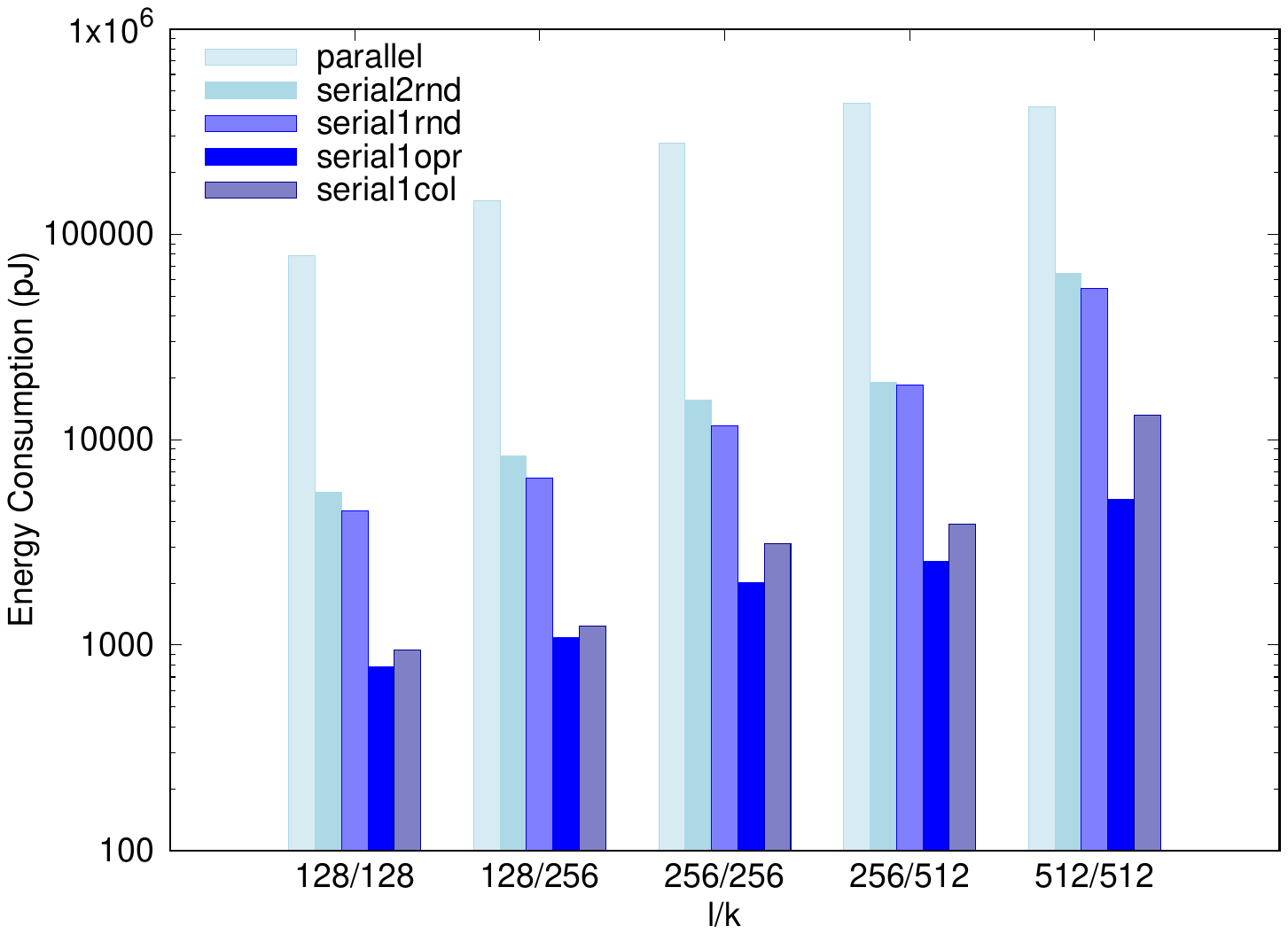}}}\
			
	\vspace*{-3.5mm}
			
	\parbox{5.8cm}{\centerline{\scriptsize (a)}}\
	\parbox{5.8cm}{\centerline{\scriptsize (b)}}\
	\parbox{5.8cm}{\centerline{\scriptsize (c)}}\
	\vspace{-3mm}
	\caption{Gate-level synthesis results of designs realizing encryption: (a)~area; (b)~latency; (c)~energy consumption.}  
	\label{fig:enc_synth}
	\vspace{-4mm}
\end{figure*}

For the encryption function, the gate-level synthesis results of designs under different architectures are given in Fig.~\ref{fig:enc_synth}, where \textit{area} is the total area in $\mu m^2$, \textit{latency} in $ns$ is the product of the number of clock cycles and the delay in the critical path, and \textit{energy consumption} in $pJ$ is the product of latency and total power dissipation. Note that the values in the \mbox{$y$-axis} are given in log scale. These designs were synthesized by Cadence Genus using a commercial 65\;nm cell library and were validated using 1000 randomly generated inputs in simulation. The switching activity data were also generated during simulation and used to determine the power dissipation. 

\begin{table}[t]
	\centering
	\caption{Number of clock cycles in Kalyna designs.}
	\vspace{-3mm}
	\footnotesize
	\begin{tabular}{|c|c|c|c|c|c|}
		\hline
		\multirow{2}{*}{Architecture} & \multicolumn{5}{c|}{$l/k$} \\
		\cline{2-6}
		& 128/128 & 128/256 & 256/256 & 256/512 & 512/512 \\ 
		\hline \hline
		parallel   & 1   & 1   & 1   & 1   & 1 \\
		serial2rnd & 5   & 7   & 7   & 9   & 9 \\
		serial1rnd & 11  & 15  & 15  & 19  & 19 \\
		serial1opr & 64  & 84  & 84  & 104 & 104 \\
		serial1col & 190 & 250 & 416 & 516 & 928 \\
		\hline
	\end{tabular}
	\label{tab:clock}
	\vspace{-6mm}
\end{table}

\begin{table*}[t]
	\centering
	\caption{Gate-level synthesis results under the \textit{serial1col} architecture.}
	\vspace{-3mm}
	\footnotesize
	\begin{tabular}{|c|c||c|c|c||c|c|c||c|c|c|}
		\hline
		\multirow{2}{*}{$k$} & \multirow{2}{*}{$l$} & \multicolumn{3}{c||}{Encryption} & \multicolumn{3}{c||}{Decryption} & \multicolumn{3}{c|}{Unified} \\ 
		\cline{3-11}
		& & area & latency & energy & area & latency & energy & area & latency & energy\\
		\hline \hline
		128 & 128 & 23662 & 792  & 944   & 35946 & 845  & 1654  & 37172  & 854  & 1705  \\
		128 & 256 & 24323 & 1042 & 1237  & 36548 & 1108 & 2185  & 37975  & 1125 & 2263  \\
		256 & 256 & 39549 & 1782 & 3109  & 55286 & 1993 & 5228  & 57673  & 1997 & 5400  \\
		256 & 512 & 41999 & 2211 & 3880  & 57553 & 2519 & 6585  & 59331  & 2500 & 6701  \\
		512 & 512 & 73966 & 4441 & 13164 & 97411 & 4347 & 17921 & 101036 & 4322 & 18298 \\		
		\hline
	\end{tabular}
	\label{tab:serial1col}
	\vspace{-4mm}
\end{table*}

\begin{table*}[t]
	\centering
	\caption{Impact of hardware reduction techniques under the \textit{serial1col} architecture.}
	\vspace{-3mm}
	\footnotesize
	\begin{tabular}{|c||c|c|c||c|c|c||c|c|c|}
		\hline
		\multirow{2}{*}{HRT} & \multicolumn{3}{c||}{Encryption} & \multicolumn{3}{c||}{Decryption} & \multicolumn{3}{c|}{Unified} \\ 
		\cline{2-10}
		& area & latency & energy & area & latency & energy & area & latency & energy\\
		\hline \hline
		KSA-OPT   & 23662 & 792  & 944  & 35946 & 845  & 1654 & 37172 & 854  & 1705 \\
		GA-NOOPT  & 24606 & 1124 & 1440 & 38036 & 1141 & 2510 & 39297 & 1141 & 2565 \\
		\hline
	\end{tabular}
	\label{tab:hrt}
	\vspace{-4mm}
\end{table*}

Observe from Fig.~\ref{fig:enc_synth} that while the \textit{serial1col} architecture leads to designs with the smallest area but the highest latency since it uses a minimum number of 64-bit operations but requires the highest number of clock cycles, the designs under the \textit{parallel} architecture have the largest area but the smallest latency since it exploits all the necessary operations but only in one clock cycle. Moreover, the \textit{serial1opr} architecture leads to designs with the smallest energy consumption since they have the smallest area and latency product while the designs under the \textit{parallel} architecture consume the highest energy. Similar results are observed on these design architectures realizing decryption and unified encryption/decryption functions.

To further analyze the hardware complexity of the Kalyna block cipher with the lowest area obtained under the \textit{serial1col} architecture, Table~\ref{tab:serial1col} presents the gate-level synthesis results of designs realizing encryption, decryption, and unified encryption/decryption.

Observe from Table~\ref{tab:serial1col} that the area of designs does not increase proportionally to the increase in the block size $l$ and key length $k$ as observed in other design architectures. This is because of the use of \mbox{64-bit} operations. However, the latency and energy consumption increase proportionally to the increase in $l$ and $k$ due to the increase in the number of clock cycles, which is a function of both the number of rounds $t$ and the number of columns in ISM $c$ (Table~\ref{tab:clock}). The area of designs realizing decryption is larger than that of designs realizing encryption since the decryption transformation needs the SPM and modulo 2 and $2^{64}$ additions and their inverses. The unified designs have hardware complexity very close to that of designs realizing decryption since common operations are shared. It is interesting to mention that while the sequential logic occupies $33\%$ of the total area when $k$ and $l$ are 128, it occupies $44\%$ of the total area when $k$ and $l$ are 512 for the Kalyna designs realizing the encryption function. Note that similar results are observed for Kalyna designs realizing the decryption and unified encryption/decryption functions.

To find the impact of hardware reduction techniques presented in Section~\ref{subsec:da} on the hardware complexity, Table~\ref{tab:hrt} presents the gate-level synthesis results of Kalyna designs under the \textit{serial1col} architecture when $l$ and $k$ are 128.

Observe from Table~\ref{tab:hrt} that the use of KSA and hardware reduction techniques leads to $3.83\%$, $29.52\%$, and $34.44\%$ reduction in area, latency, and energy consumption, respectively, on the design realizing encryption when compared to the design using GA and substitution and mix column operations without optimizations. These values are obtained as $5.49\%$ ($5.40\%$), $25.94\%$ ($25.14\%$), and $34.10\%$ ($33.52\%$) in area, latency, and energy consumption, respectively, on the design realizing decryption (encryption/decryption). Observe that these techniques have more impact on latency and energy consumption than area since they reduce the delay significantly. Note that similar results are observed on designs with other $l$ and $k$ values.

The maximum frequency of the Kalyna design realizing encryption under the \textit{serial1col} architecture when $l$ and $k$ are 128 is determined by changing the delay constraint in a binary search manner. Table~\ref{tab:mfreq} presents the gate-level synthesis results of Kalyna designs found in this process, where \textit{delay} and \textit{power} denote the delay in the critical path in $ns$ and total power dissipation in $mW$. 

\begin{table}[t]
	\centering
	\caption{Impact of a delay constraint under the \textit{serial1col} architecture.}
	\vspace{-3mm}
	\footnotesize
	\begin{tabular}{|c|c|c|c|c|}
		\hline
		area & delay & latency & power & energy \\ 
		\hline \hline
		23662 & 4.171 & 792 & 1.191 & 944 \\
		24103 & 2.053 & 390 & 1.143 & 446 \\
		24582 & 1.491 & 283 & 1.194 & 338 \\
		28479 & 1.228 & 233 & 1.291 & 301 \\
		31756 & 1.083 & 205 & 1.514 & 312 \\
		35010 & 1.029 & 195 & 1.671 & 327 \\
		\hline
	\end{tabular}
	\label{tab:mfreq}
	\vspace{-4mm}
\end{table}

\begin{table}[t]
	\centering
	\caption{Gate-level synthesis results of AES and Kalyna.}
	\vspace{-3mm}
	\footnotesize
	\begin{tabular}{|c|c|c|c|}
		\hline
		Cipher & area & latency & energy \\ 
		\hline \hline
		AES    & 187962 & 19673 & 9423 \\
		Kalyna & 400166 & 44462 & 78748 \\
		\hline
	\end{tabular}
	\label{tab:aes}
	\vspace{-6mm}
\end{table}

Observe from Table~\ref{tab:mfreq} that designs with different area, latency, and energy consumption values can be obtained by changing the delay constraint in the design requirement. The design latency can be reduced by $4.05\times$, leading to a $2.88\times$ decrease in energy consumption but a $1.47\times$ increase in area. In this case, the clock frequency reaches up to 971 MHz, which is a rather high frequency of operation for a 65nm accelerator.

In order to compare the hardware complexity of the Kalyna block cipher with that of the well-known AES, the parallel design architecture realizing the encryption function when $l$ and $k$ are 128 is considered, since there exist various hardware-efficient AES implementations and the parallel design architecture, implemented as a full combinational circuit, executes all the operations in one clock cycle. Table~\ref{tab:aes} presents the gate-level synthesis results of Kalyna and AES.

Observe from Table~\ref{tab:aes} that the hardware complexity of Kalyna is significantly larger than that of AES, i.e., $2.12\times$ in area, $2.26\times$ in latency, and $8.35\times$ in energy consumption. This is simply because of Kalyna's requirement to have four different S-boxes, modulo $2^{64}$ addition operation, and a complex mix column operation and key schedule.

\subsection{Security Countermeasures}

\begin{figure*}[t]
	\centering
	\vspace{-6mm}
	\parbox{5.8cm}{\centerline{\includegraphics[width=6.5cm]{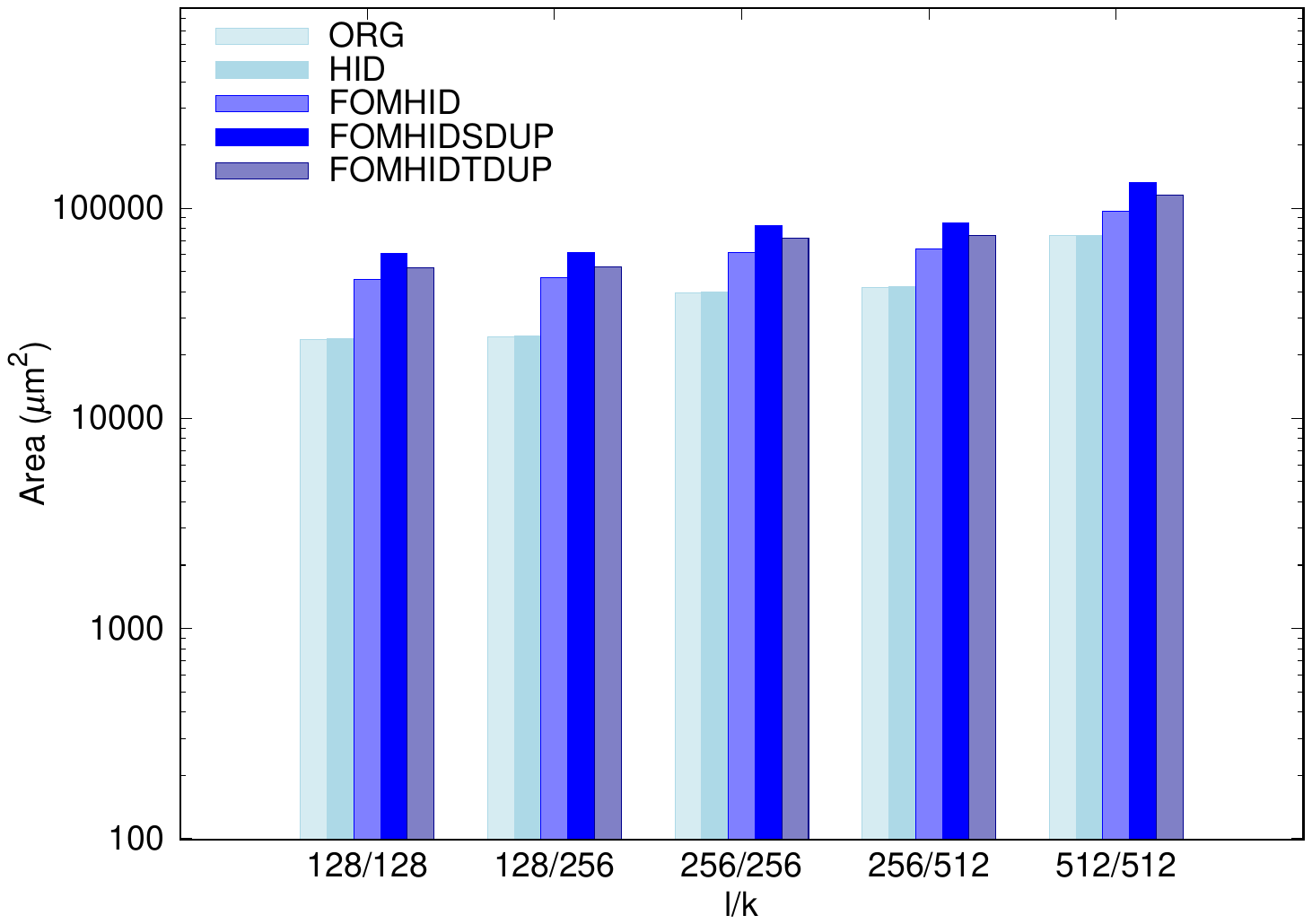}}}\
	\parbox{5.8cm}{\centerline{\includegraphics[width=6.5cm]{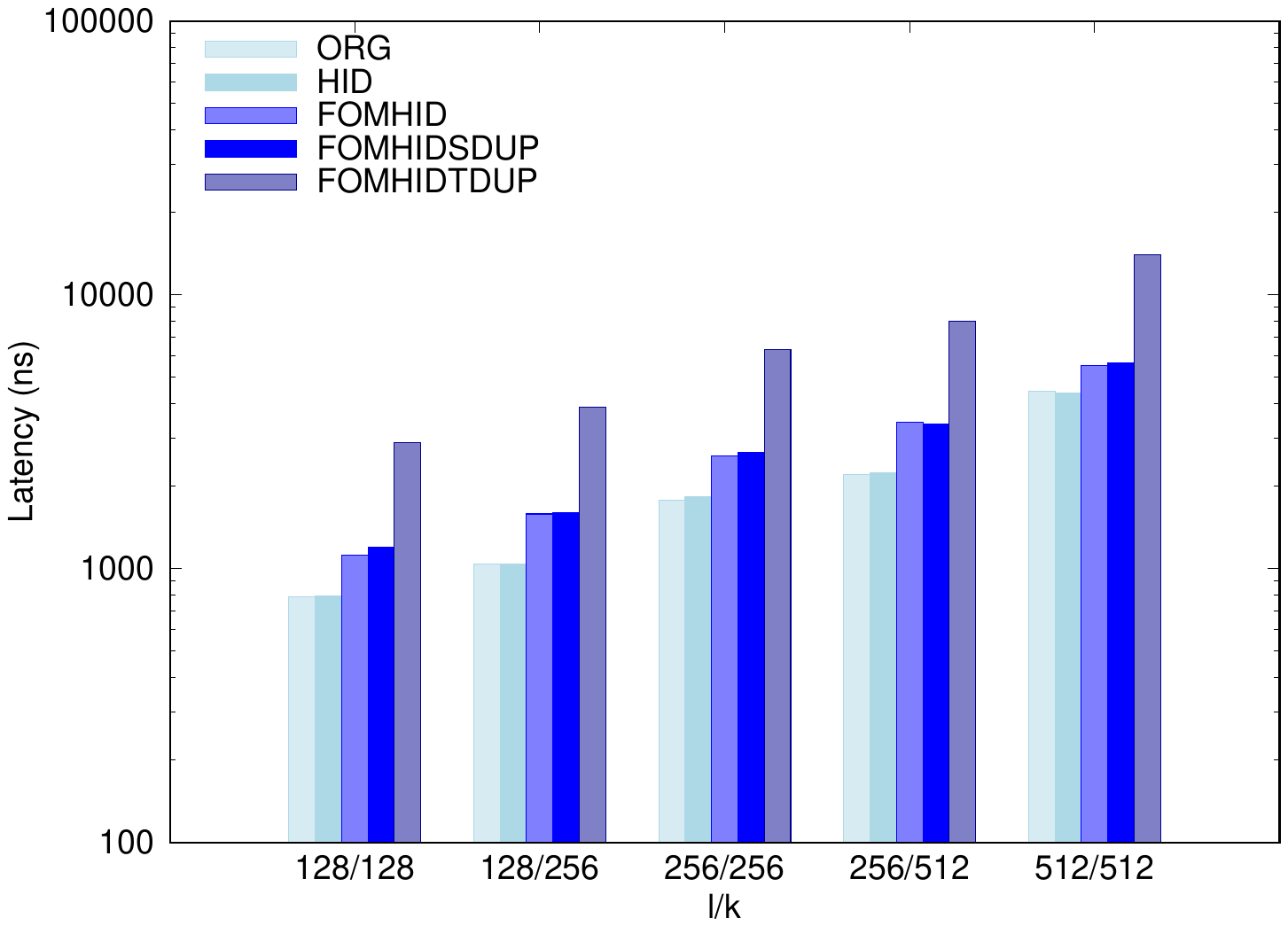}}}\
	\parbox{5.8cm}{\centerline{\includegraphics[width=6.5cm]{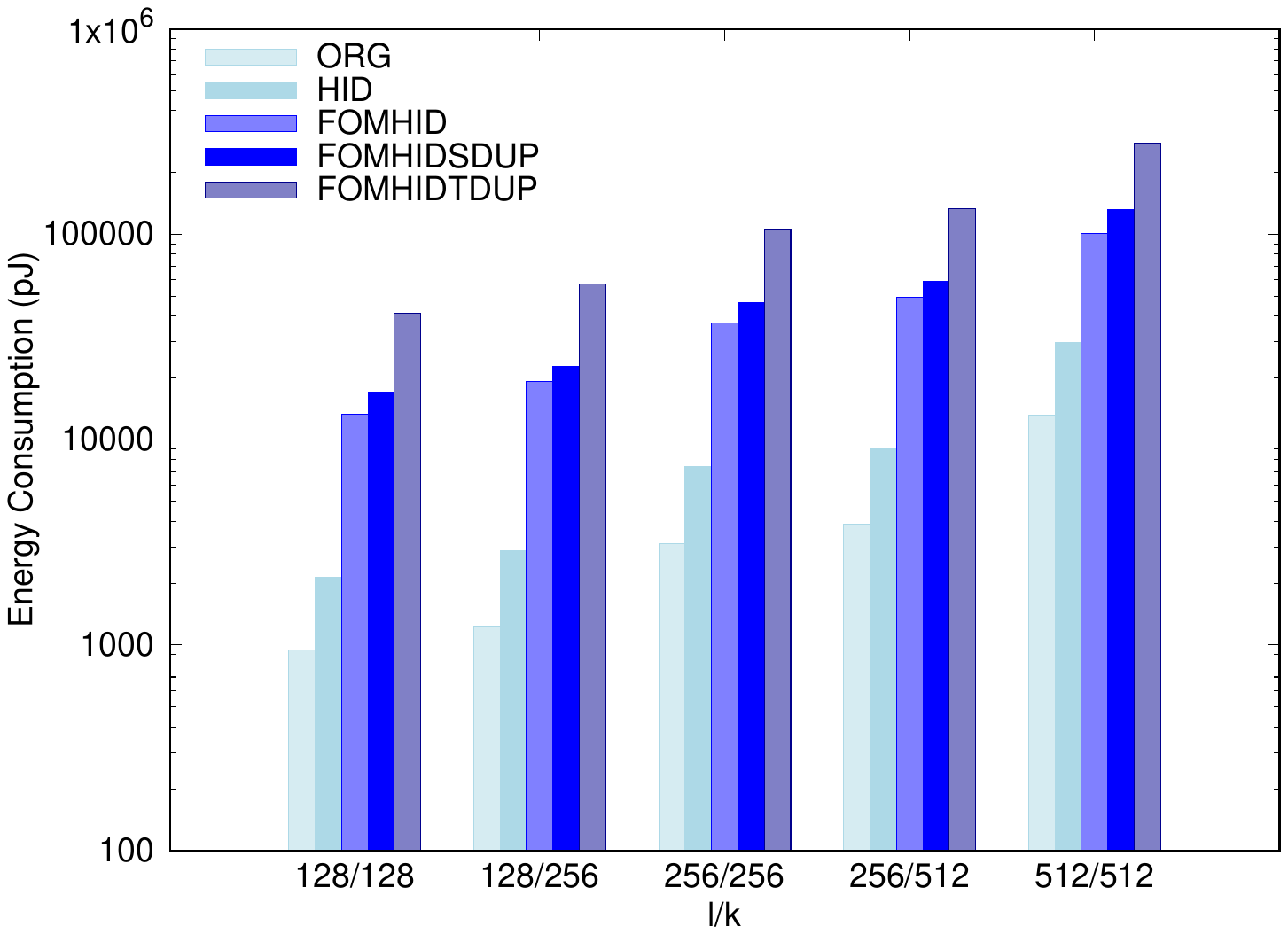}}}\
			
	\vspace*{-3.5mm}
			
	\parbox{5.8cm}{\centerline{\scriptsize (a)}}\
	\parbox{5.8cm}{\centerline{\scriptsize (b)}}\
	\parbox{5.8cm}{\centerline{\scriptsize (c)}}\
	\vspace{-2mm}
	\caption{Gate-level synthesis results of designs without and with countermeasures realizing encryption: (a)~area; (b)~latency; (c)~energy consumption.}  
	\label{fig:secenc_synth}
	\vspace{-4mm}
\end{figure*}

The countermeasures of hiding (HID), first-order masking and hiding (FOMHID), first-order masking, hiding, and spatial duplication (FOMHIDSDUP), and first-order masking, hiding, and temporal duplication (FOMHIDTDUP) are realized on the \textit{serial1col} architecture, since it leads to designs with the smallest area. Fig.~\ref{fig:secenc_synth} presents the gate-level synthesis results of these designs in addition to the original design without any countermeasures (ORG) under the same design architecture realizing the encryption function.

\begin{table*}[t]
	\centering
	\caption{Gate-level synthesis results of designs with the FOMHIDTDUP techniques.}
	\vspace{-3mm}
	\footnotesize
	\begin{tabular}{|c|c||c|c|c||c|c|c||c|c|c|}
		\hline
		\multirow{2}{*}{$k$} & \multirow{2}{*}{$l$} & \multicolumn{3}{c||}{Encryption} & \multicolumn{3}{c||}{Decryption} & \multicolumn{3}{c|}{Unified} \\ 
		\cline{3-11}
		& & area & latency & energy & area & latency & energy & area & latency & energy\\
		\hline \hline
		128 & 128 & 52034  & 2888  & 41174  & 80607  & 2397  & 83821  & 89928  & 2473  & 105831 \\
		128 & 256 & 52790  & 3876  & 57400  & 75468  & 3309  & 95925  & 85946  & 3264  & 122187 \\
		256 & 256 & 72248  & 6305  & 106340 & 111402 & 5989  & 236774 & 114390 & 5597  & 240024 \\
		256 & 512 & 73836  & 8021  & 133474 & 108953 & 7218  & 262354 & 117045 & 7484  & 325729 \\
		512 & 512 & 115216 & 13984 & 278790 & 162296 & 12529 & 633760 & 166601 & 12617 & 626716 \\
		\hline
	\end{tabular}
	\label{tab:fomhidtdup_synth}
	\vspace{-4mm}
\end{table*}

\begin{table}[t]
	\centering
	\caption{Impact of a delay constraint under the \textit{serial1col} architecture with the FOMHIDTDUP countermeasures.}
	\vspace{-3mm}
	\footnotesize
	\begin{tabular}{|c|c|c|c|c|}
		\hline
		area & delay & latency & power & energy \\ 
		\hline \hline
		52034 & 7.600 & 2888 & 14.257 & 41174 \\
		52234 & 3.936 & 1495 & 14.597 & 21832 \\
		61393 & 1.921 & 729  & 24.944 & 18209 \\
		75933 & 1.641 & 623  & 32.266 & 20120 \\
		87492 & 1.548 & 588  & 39.646 & 23321 \\
		\hline
	\end{tabular}
	\label{tab:secenc_mfreq}
	\vspace{-6mm}
\end{table}

Observe from Fig.~\ref{fig:secenc_synth} that hiding (HID) can be applied with a slight increase in hardware complexity with respect to the original design since available resources are used. Note that the power dissipation is increased up to $2.3\times$ in this case since several operations are executed at the same time. The use of first-order masking in addition to hiding (FOMHID) increases area, latency, and energy consumption up to $1.9\times$, $1.5\times$, and $15.5\times$, respectively, with respect to the original design. This is simply because of the addition of extra logic required by Boolean masking. The spatial duplication in addition to the countermeasures against the SCA attacks (FOMHIDSDUP) increases the area, latency, and energy consumption up to $2.5\times$, $1.5\times$, and $18.4\times$, respectively, with respect to the original design due to the inherent duplication of operations. The temporal duplication, in addition to the countermeasures SCA attacks (FOMHIDTDUP), increases the area, latency, and energy consumption up to $2.2\times$, $3.7\times$, and $46.4\times$, respectively, with respect to the original design. Observe that the area overhead in designs using the FOMHIDTDUP techniques is smaller than that in designs using the FOMHIDSDUP techniques since the temporal duplication re-uses the available operations, but doubles the number of clock cycles, increasing the latency and energy consumption. 


Table~\ref{tab:fomhidtdup_synth} presents the gate-level synthesis results of the Kalyna designs with the FOMHIDTDUP countermeasures under the \textit{serial1col} architecture realizing the encryption, decryption, and unified encryption/decryption functions. 

Observe from Table~\ref{tab:fomhidtdup_synth} that the area and energy consumption of designs realizing the decryption function increase up to $1.5\times$  and $2.2\times$, respectively, when compared to those of designs realizing the encryption function since they also include the modulo $2^{64}$ subtraction and inverse SPM operations. Moreover, the hardware complexity of designs realizing the unified encryption/decryption functions is increased slightly with respect to those realizing the decryption function since they use the same operations.

Table~\ref{tab:secenc_mfreq} presents the gate-level synthesis results of Kalyna designs with the FOMHIDTDUP countermeasures under the \textit{serial1col} architecture realizing the encryption function when $l$ and $k$ are 128, while the maximum frequency is explored by changing the delay constraint in a binary search manner. 

Observe from Table~\ref{tab:secenc_mfreq} that the design latency can be reduced by $4.9\times$, leading to a $1.76\times$ decrease in energy consumption, but a $1.68\times$ increase in area. In this case, the clock frequency reaches up to 645\;MHz. Note that the maximum clock frequency decreases around $33\%$ when compared to the original design without any countermeasures (Table~\ref{tab:mfreq}).

\subsection{Physical Designs}
\label{subsec:backend}

To explore the hardware complexity of physical designs of Kalyna, the ones under the \textit{serial1col} design architecture without and with FOMHIDTDUP countermeasures realizing the unified encryption/decryption function when $l$ and $k$ are 128, are considered. For the physical synthesis, Cadence Innovus was used along with a commercial 65\;nm PDK.

\begin{figure}[t]
	\centering
	\scalebox{.74}{
		\begin{tikzpicture}
			\hspace{-7mm}
			\node[anchor=south west,inner sep=0] (image1) at (0.5,0) {\includegraphics[width=0.41\columnwidth, height=0.42\columnwidth]{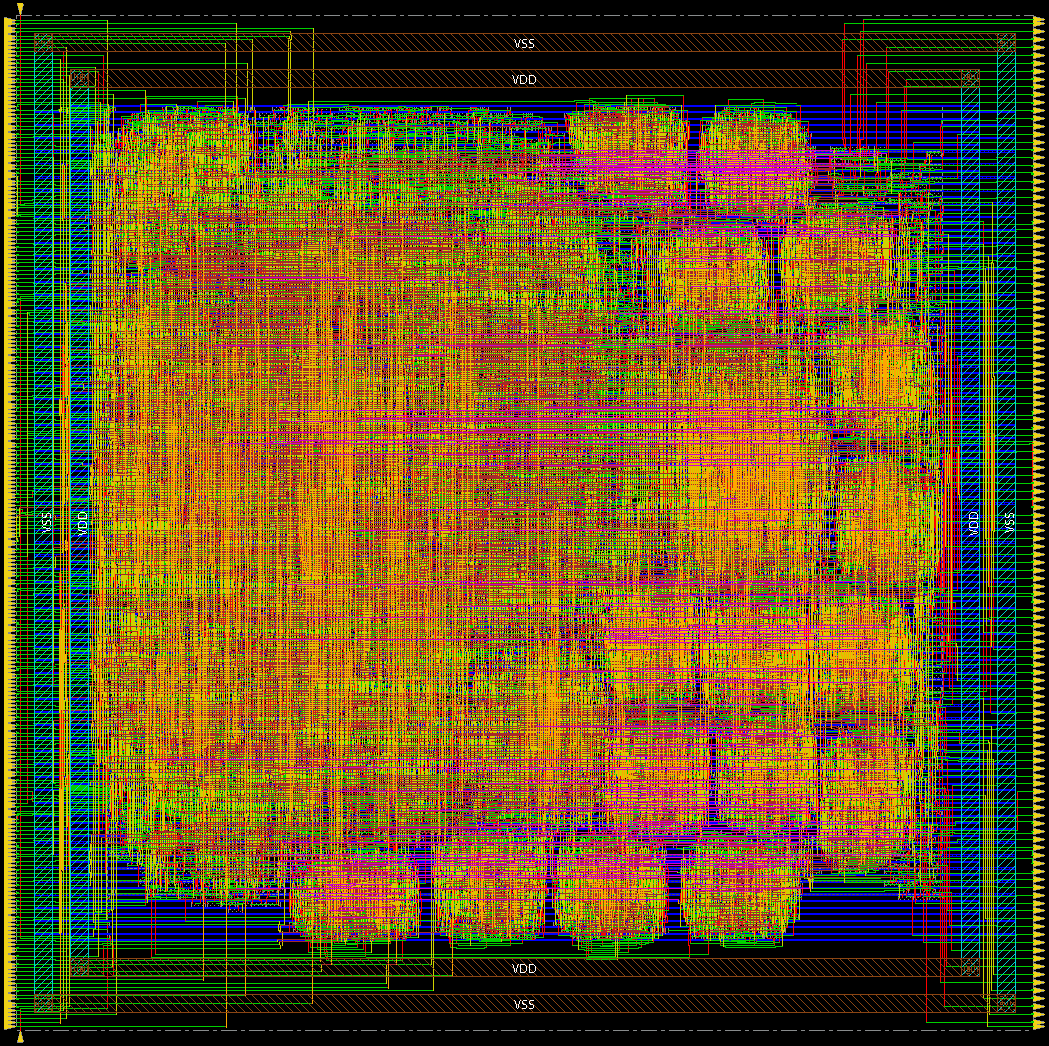}};
			\draw [<->](0.40 , 0) -- (0.40 , 3.7) node[left, font=\small, rotate=90] at (0.1,2.5) {\: 280.4 \si{\micro\meter}};
			\draw [<->](0.5 , -.10) -- (4.15 , -.10) node[midway, below, font=\small] {\: 280.8 \si{\micro\meter}};
			\hspace{10mm}
			\node[anchor=south west,inner sep=0] (image2) at (5,0) {\includegraphics[width=0.60\columnwidth, height=0.60\columnwidth]{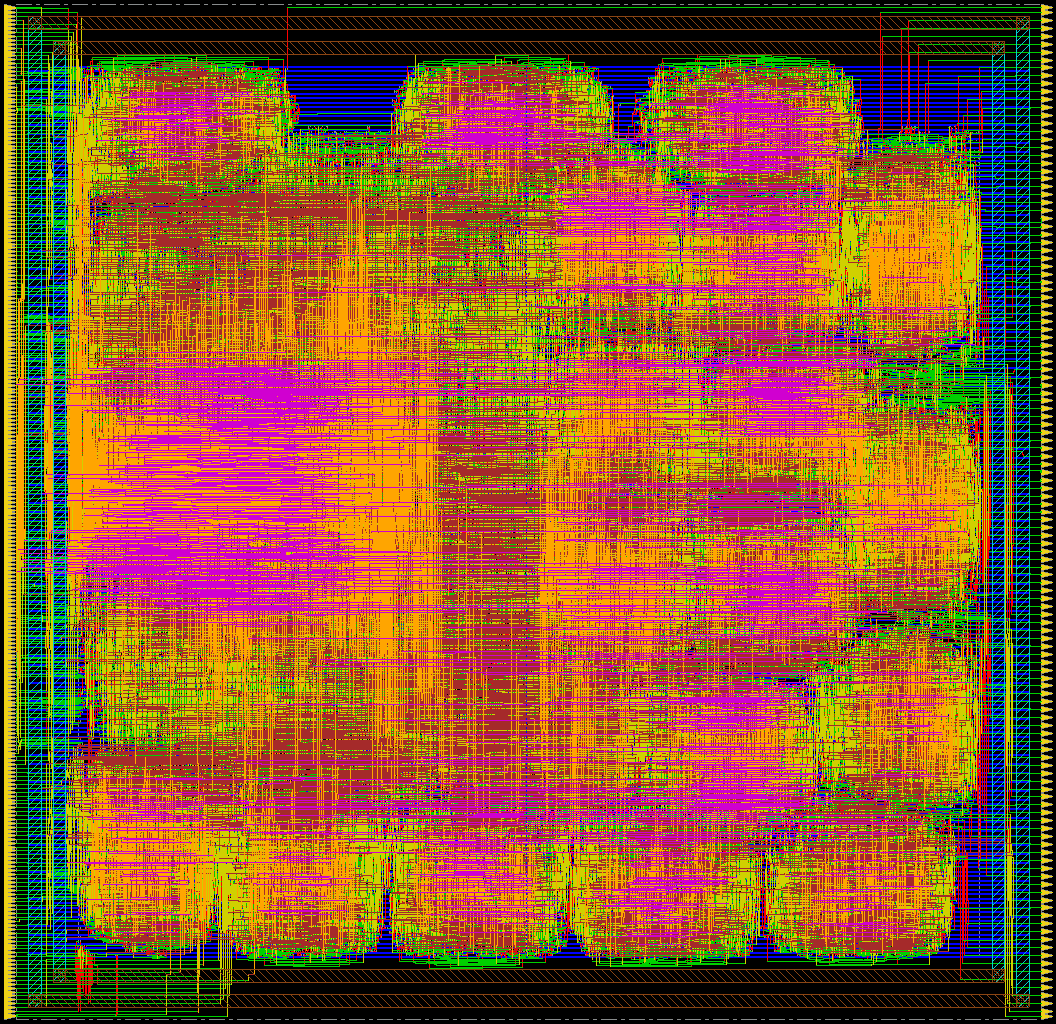}};
			\draw [<->](4.9 , 0) -- (4.9 , 5.3) node[left, font=\small, rotate=90] at (4.6,3.5) {\: 406.4 \si{\micro\meter}};
			\draw [<->](5 , -.10) -- (10.35 , -.10) node[midway, below, font=\small] {\: 410.6 \si{\micro\meter}};
			\node[font=\normalsize] at (1.4,-1.0) {(a)};
			\node[font=\normalsize] at (7.8,-1.0) {(b)};
		\end{tikzpicture} 
	}
	\vspace{-4mm}
	\caption{Layouts with routed view of the Kalyna block cipher: (a)~without countermeasures; (b)~with the FOMHIDTDUP countermeasures.}
	\vspace{-2mm}
	\label{fig:layout_routed}
\end{figure}
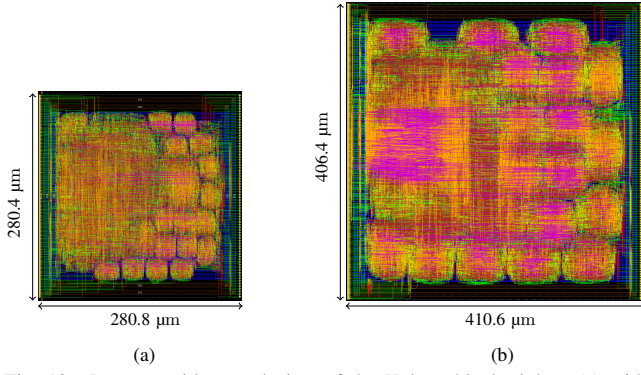

\begin{figure}[t]
	\centering
	\scalebox{.74}{
		\begin{tikzpicture}
			\hspace{-4mm}
			\node[anchor=south west,inner sep=0] (image1) at (0.5,0) {\includegraphics[width=0.41\columnwidth, height=0.42\columnwidth]{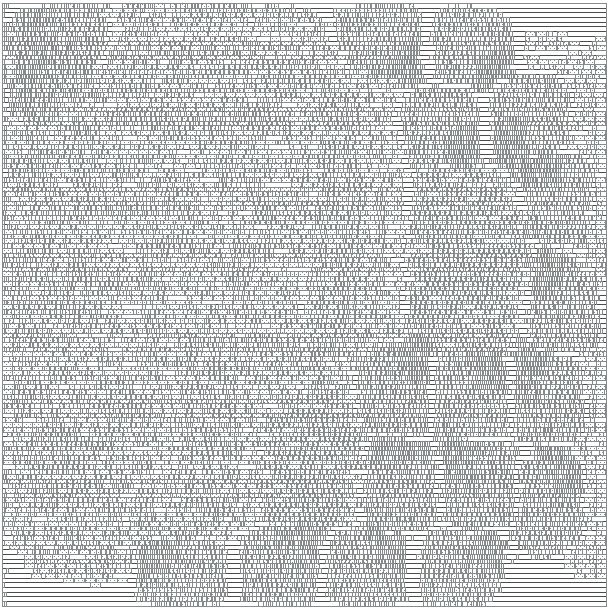}};
			\hspace{10mm}
			\node[anchor=south west,inner sep=0] (image2) at (5,0) {\includegraphics[width=0.60\columnwidth, height=0.60\columnwidth]{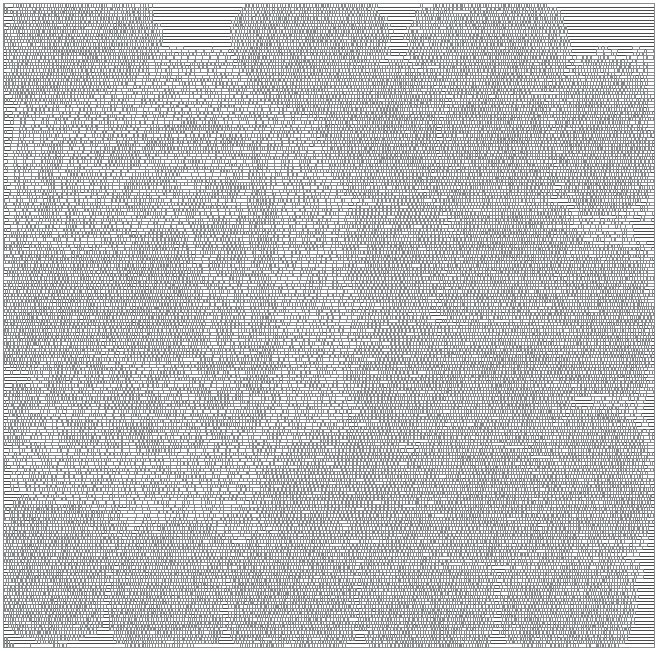}};
			\vspace{-12mm}
			\node[font=\normalsize] at (1.4,-1.0) {(a)};
			\node[font=\normalsize] at (7.8,-1.0) {(b)};
		\end{tikzpicture} 
	}
	\vspace{-4mm}
	\caption{Layouts with placement view of the Kalyna block cipher: (a)~without countermeasures; (b)~with the FOMHIDTDUP countermeasures.}
	\vspace{-6mm}
	\label{fig:layout_placed}
\end{figure}

Fig.~\ref{fig:layout_routed} presents layouts of the selected Kalyna designs under their routed views. To ensure a fair comparison between these two designs, the same initial placement density, i.e., $70\%$, was used during floorplanning. This consistent setup results in noticeable differences in the overall layout dimensions, effectively illustrating the area disparity between these designs. These layouts highlight the impact of the FOMHIDTDUP countermeasures on the silicon area, indicating the trade-off between security enhancements and hardware overhead, visually. Observe from Tables~\ref{tab:serial1col} and~\ref{tab:fomhidtdup_synth} that the area of the design with countermeasures is $2.4\times$ larger than that of the design without countermeasures, while the physical design with countermeasures is $2.1\times$ larger than that of the design without countermeasures. Fig.~\ref{fig:layout_placed} shows the placement density of the physical designs given in Fig.~\ref{fig:layout_routed}. Although the overall layout dimensions differ due to variations in area, the uniformity in the placement cell density is clearly visible in this figure. 

Table~\ref{tab:pdr} presents the hardware complexity of these physical designs. It is important to note here that the \textit{total area} is not the die size, instead it represents the sum of the area of all standard cells. Observe that these values are very close to those given in Tables~\ref{tab:serial1col} and~\ref{tab:fomhidtdup_synth}. In this table, \textit{slack} denotes the difference between the required clock period of 10\;ns and the delay in the critical path. Note that these designs have positive slacks, confirming that they respect the given clock frequency. This table also presents various types of power dissipation values, indicating that the Kalyna design with countermeasures consumes $5.5\times$ more power than the design without countermeasures.

\begin{table}[t]
	\centering
	\caption{Physical design results of Kalyna realizing unified encryption/decryption.}
	\vspace{-3mm}
	\footnotesize
	\begin{tabular}{|@{\hskip3pt}l@{\hskip3pt}|@{\hskip3pt}c@{\hskip3pt}|@{\hskip3pt}c@{\hskip3pt}|}
		\hline
		Parameters & without countermeasures & with countermeasures \\ 
		\hline \hline
		Total area ($\mu m^2$)  & 38109 & 91920 \\
		Slack (ns)              & 3.526 & 1.993 \\
		Internal Power (mW)     & 2.692 & 10.947 \\
		Switching Power (mW)    & 2.239 & 16.279 \\
		Leakage Power (mW)      & 0.002 & 0.004 \\
		Total Power (mW)        & 4.933 & 27.230 \\
		\hline
	\end{tabular}
	\label{tab:pdr}
	\vspace{-6mm}
\end{table}

\subsection{Test Vector Leakage Assessment}
\label{subsec:tvla}

In order to understand the leakage properties of our Kalyna implementations, a TVLA flow similar to~\cite{kiaei22} was developed. This process includes four main steps: (i)~\textit{logic synthesis}, where the gate-level netlist and timing constraints are obtained by Cadence Genus; (ii)~\textit{gate-level simulation}, where the two groups of test vectors are applied and toggle values of all nets in the gate-level netlist are obtained in the value change dump (VCD) format by Cadence Xcelium; (iii)~\textit{physical design}, where the gate-level netlist is placed and routed, leading to a layout without any violations, and the power traces are generated for each test vector in two groups by Cadence Innovus using the related VCD files and timing constraints; (iv)~\textit{leakage assessment}, where the power leakage is computed using T-test as follows:
\begin{equation}
	T = \frac{\mu(S_1)-\mu(S_2)}{\sqrt{\frac{\sigma^2(S_1)}{|S_1|} + \frac{\sigma^2(S_2)}{|S_2|}}}
	\label{eqn:ttest}
\end{equation}
where $S_1$ and $S_2$ are the power measurements in the first and second groups at a time sample, respectively and $\mu(M)$, $\sigma^2(M)$, and $|M|$ denote the average, sample variance, and number of measurements $M$, respectively. If the absolute value of the T-test is greater than 4.5, it is confirmed that the design is leaking with $95\%$ confidence. At the \textit{backend design} stage, all possible corner cases of the design library are considered, and the power traces are sampled at each 1\;ns when the clock period is 100\;ns. At the \textit{leakage assessment} stage, the power traces obtained at each time sample on both first and second groups are checked if they are normally distributed using the Jarque-Bera test~\cite{jarque87}.

\begin{figure*}[t]
	\centering
	\vspace{-6mm}
	\parbox{5.8cm}{\centerline{\includegraphics[width=6.5cm]{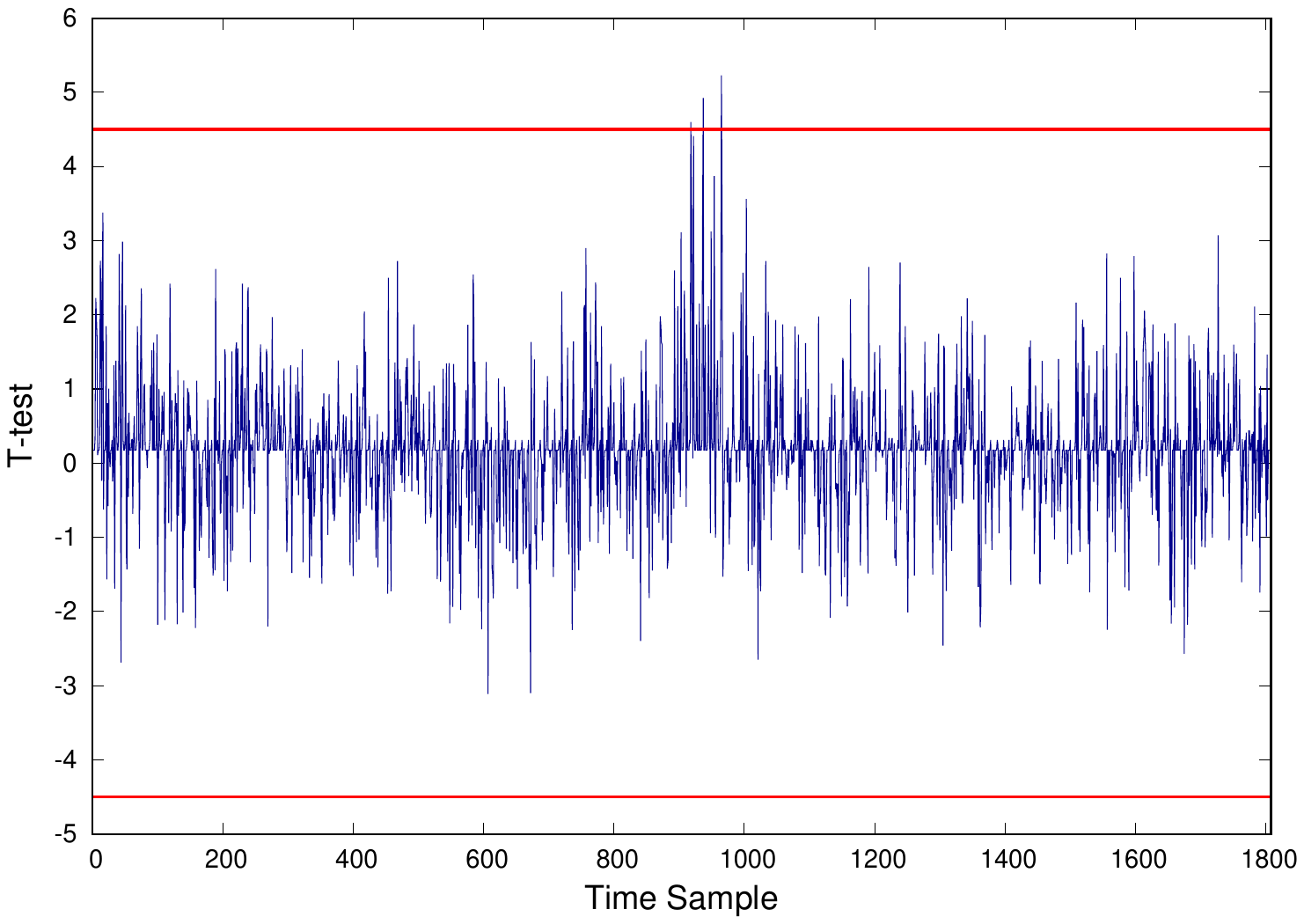}}}\
	\parbox{5.8cm}{\centerline{\includegraphics[width=6.5cm]{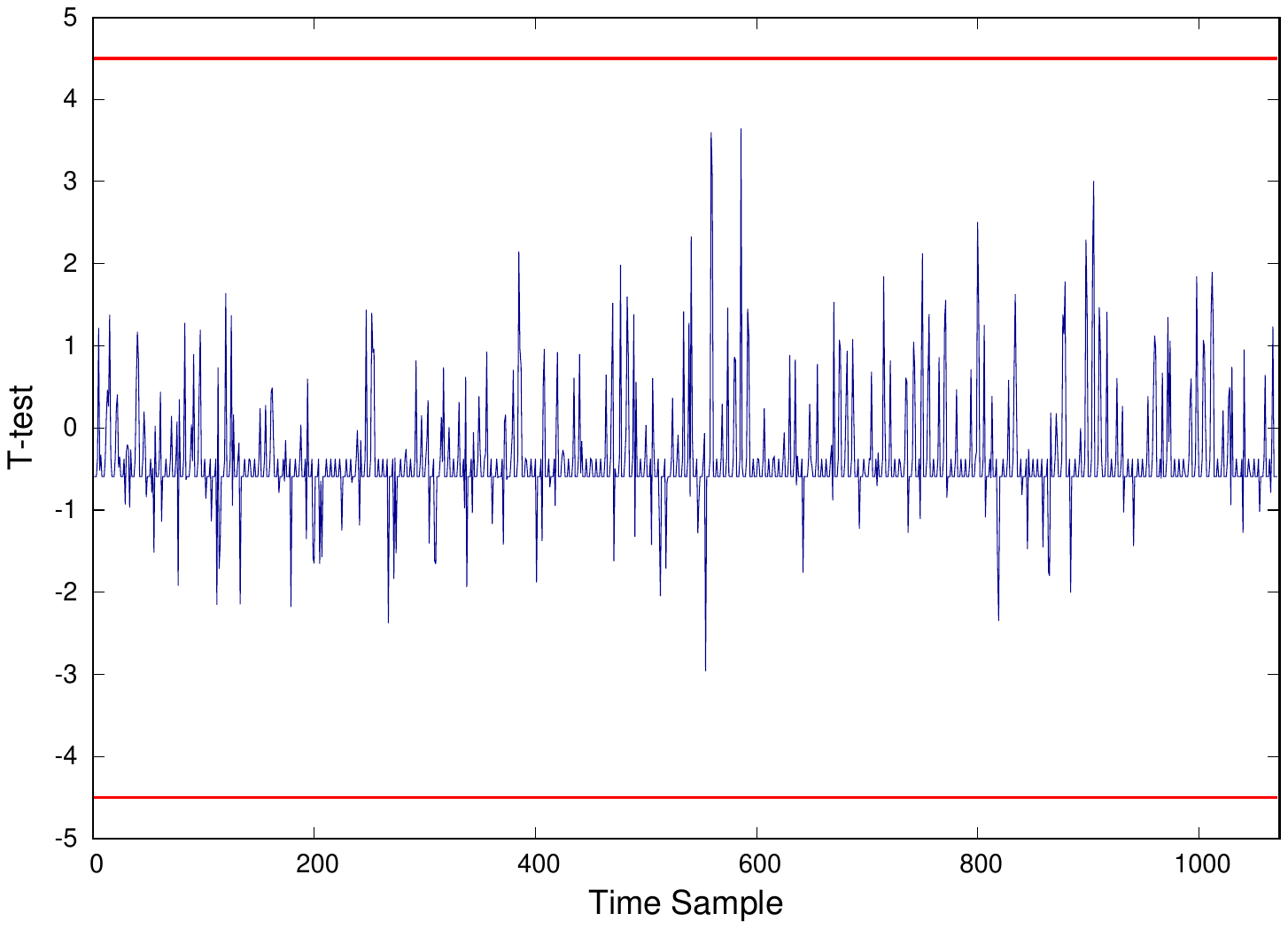}}}\
	\parbox{5.8cm}{\centerline{\includegraphics[width=6.5cm]{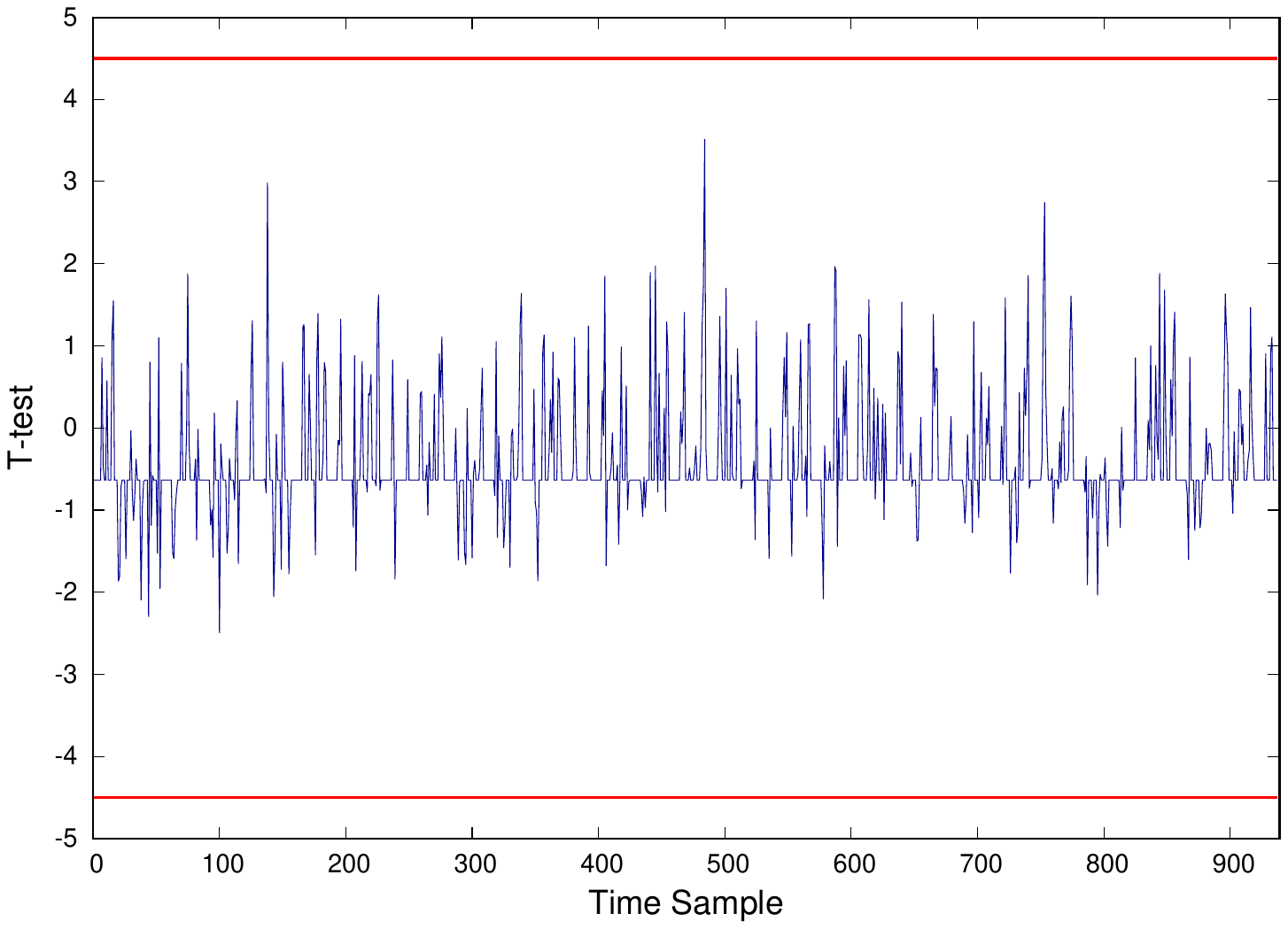}}}\
	\\
	\vspace*{-3.5mm}
	\parbox{5.8cm}{\centerline{\scriptsize (a)}}\
	\parbox{5.8cm}{\centerline{\scriptsize (b)}}\
	\parbox{5.8cm}{\centerline{\scriptsize (c)}}\

    \parbox{6.3cm}{\centerline{\includegraphics[width=6.5cm]{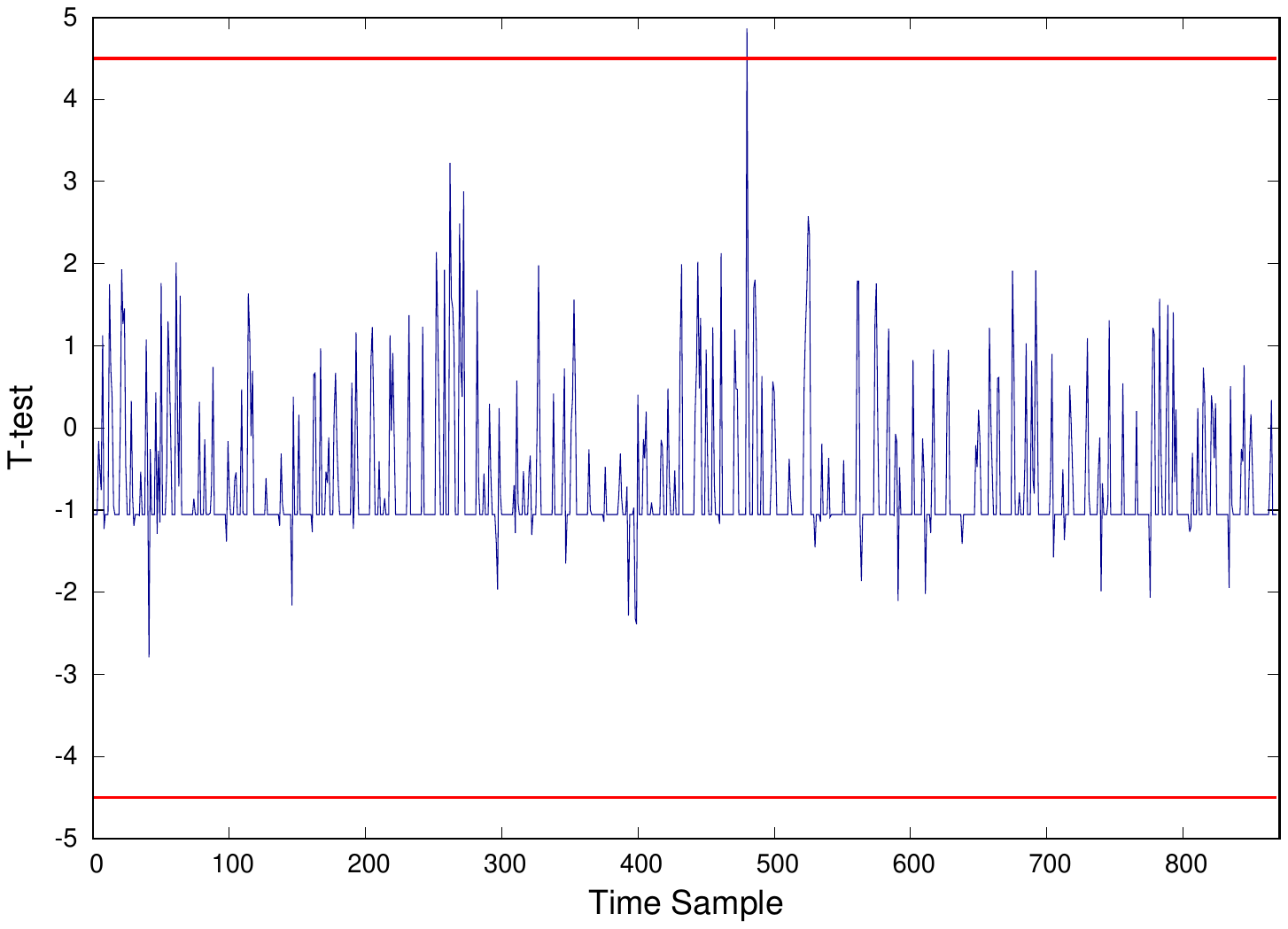}}}\
	\parbox{6.3cm}{\centerline{\includegraphics[width=6.5cm]{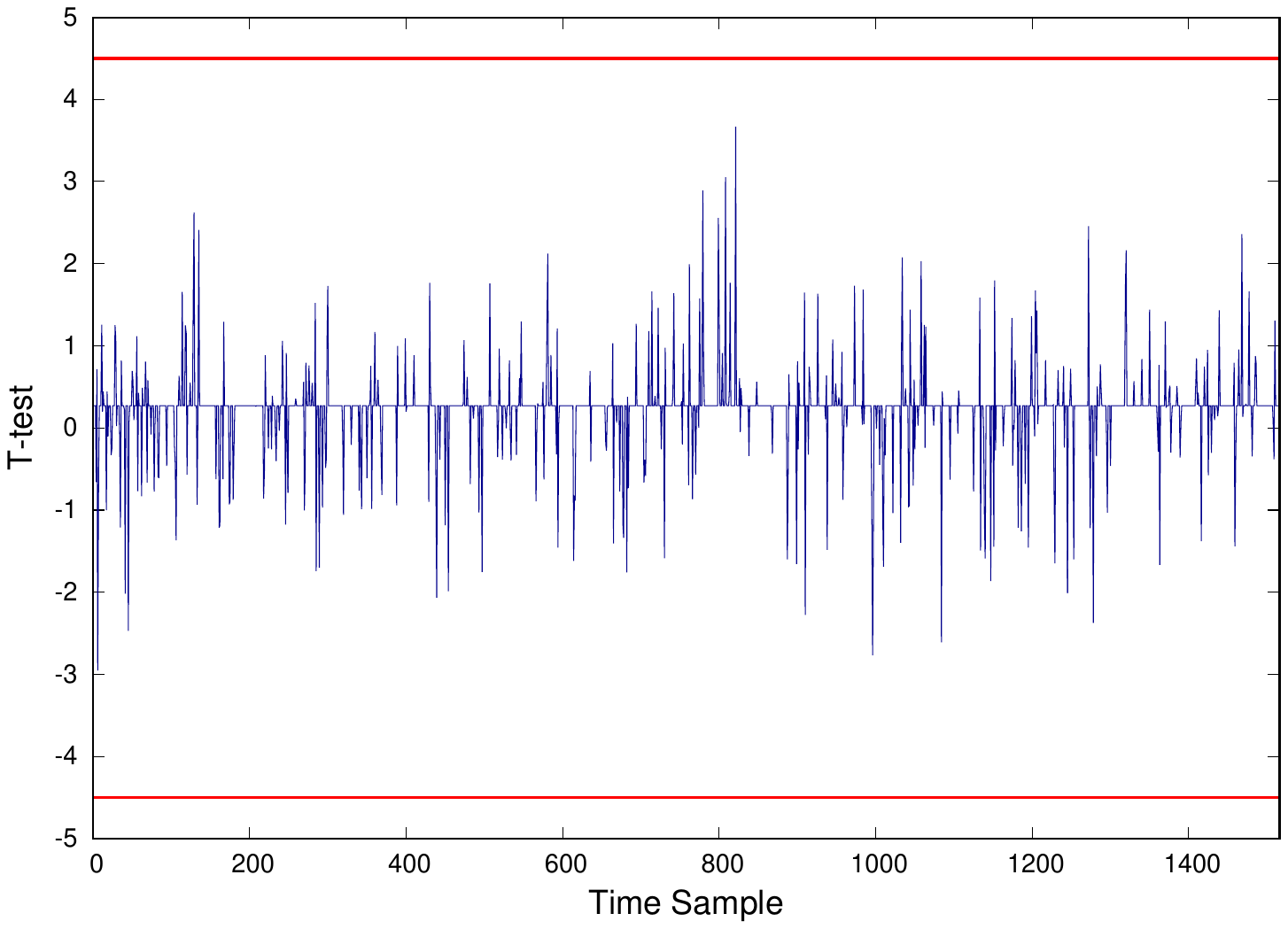}}}\
    \\
	\vspace*{-3.5mm}
	\parbox{5.8cm}{\centerline{\scriptsize (d)}}\
	\parbox{5.8cm}{\centerline{\scriptsize (e)}}\
	\vspace{-2mm}
	\caption{TVLA results: (a)~design without countermeasures; designs with countermeasures: (b)~HID; (c)~FOMHID; (d)~FOMHIDSDUP; (e)~FOMHIDTDUP.}  
	\label{fig:tvla_enc}
	\vspace{-2mm}
\end{figure*}

In addition to all designs with countermeasures under the \textit{serial1col} architecture, the original design without any countermeasures is considered for TVLA. Two groups of test vectors, (i)~random and (ii)~specific, are obtained, where the first group includes 10000 randomly generated plaintexts and keys, and the second group includes 10000 specific plaintexts and keys that generate at least 1 byte of zeros at the output of each operation at the fourth round. Fig.~\ref{fig:tvla_enc} presents the \mbox{T-test} values of designs without and with countermeasures realizing the encryption function when $bs$ and $kl$ are 128, where the red lines denote the threshold. 

Observe from Fig.~\ref{fig:tvla_enc} that while the original design leaks power, passing over the threshold, hiding (HID) and the first-order masking and hiding (FOMHID) countermeasures reduce the leakage by adding randomization in the execution of operations. It is interesting to note that the hiding countermeasure can reduce the power leakage with a slight increase in hardware complexity, as shown in Fig.~\ref{fig:secenc_synth}. However, the spatial duplication with the countermeasures against FIA and SCA attacks (FOMHIDSDUP) increases the power leakage, making the block cipher vulnerable to the SCA attacks. This is simply because duplicating the operations amplifies the power dissipation at a certain time. This drawback is overcome by the temporal duplication in designs using the countermeasures against the SCA attacks (FOMHIDTDUP). 

    \section{Silicon Validation}
\label{sec:silicon}

This section discusses the test chip fabricated in a 65nm technology that contains a subset of the Kalyna implementations described in Section~\ref{sec:architectures}. The test chip itself and its measurement setup are described, along with measurements pertaining to the silicon implementations being evaluated. For comparison purposes, we also consider a 32-bit datapath AES encryption-only implementation with on-the-fly key schedule expansion~\cite{aes_secworks}. 

\begin{figure}[t]
    \centering
    \includegraphics[angle=270,width=0.85\linewidth]{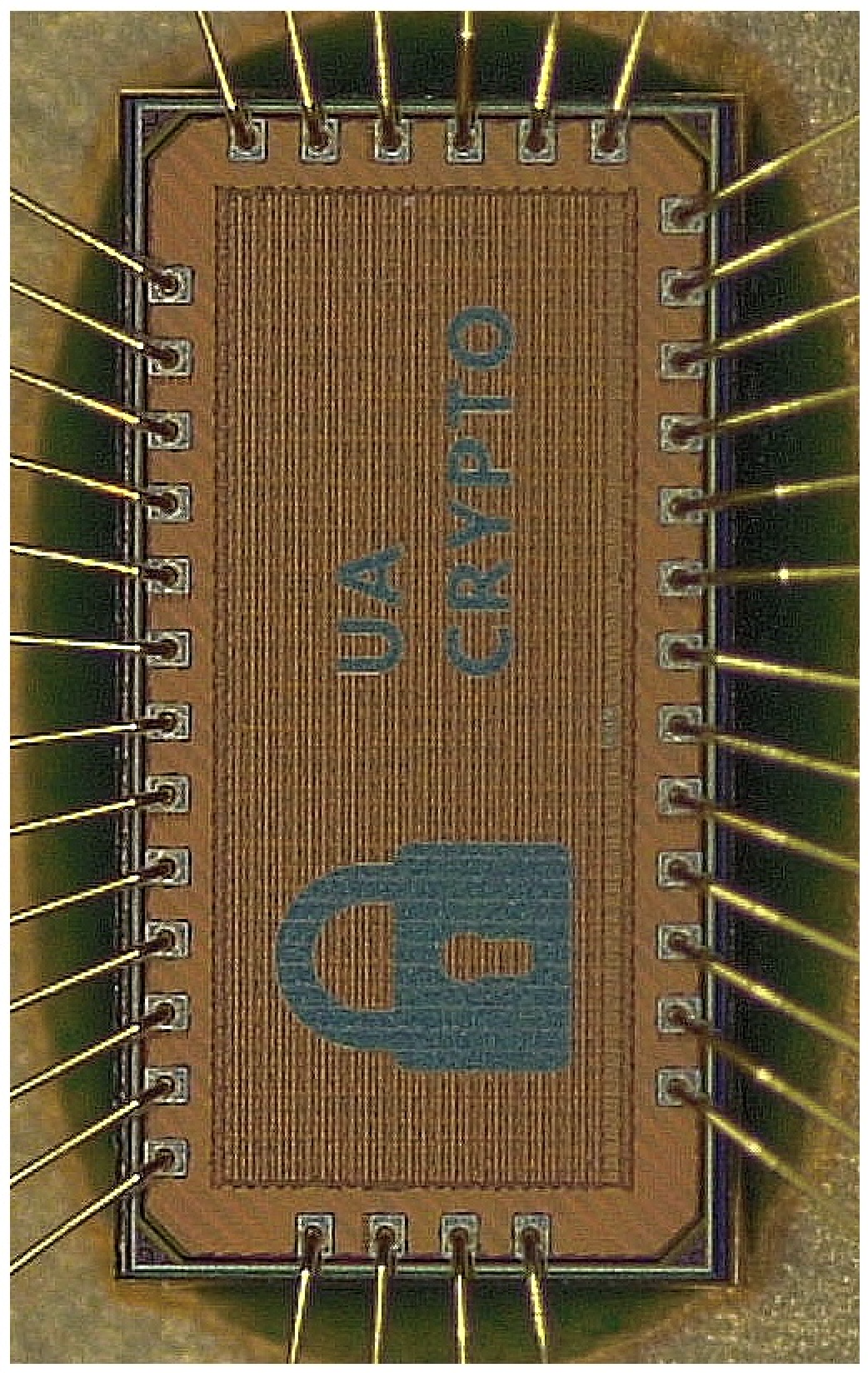}
    \caption{Test chip die.}
    \label{fig:chip_pic}
    \vspace*{-6mm}
\end{figure}

\begin{figure}[t]
    \centering
    \includegraphics[width=1\linewidth]{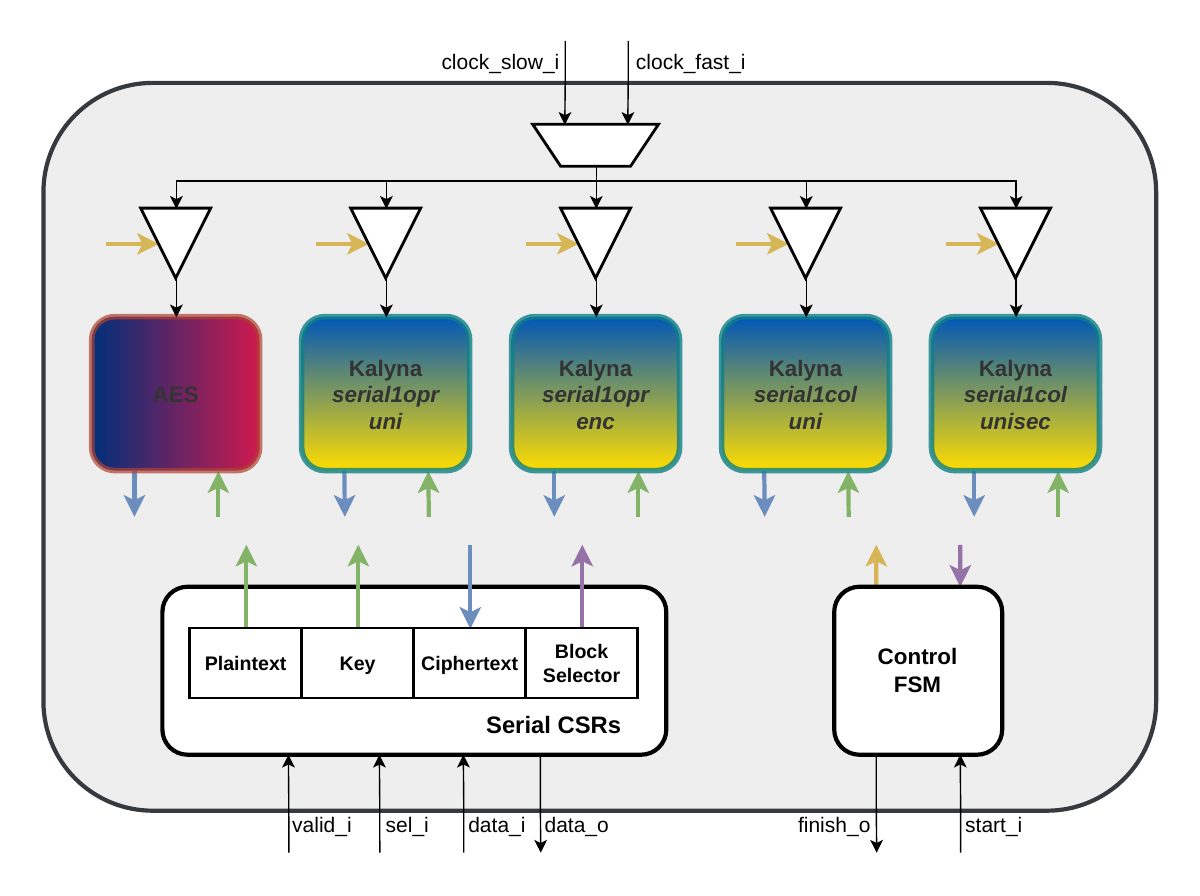}
    \vspace*{-6mm}
    \caption{Test chip top-level block diagram.}
    \label{fig:chip_block_diagram}
    \vspace*{-2mm}
\end{figure}

The test chip is depicted in Fig.~\ref{fig:chip_pic}, showing a 1 mm by 2 mm die, which contains the Kalyna and AES implementations. The Kalyna designs under the \textit{serial1opr} architecture realizing encryption and unified encryption/decryption functions, called \textit{serial1opr\_enc} and \textit{serial1opr\_uni}, respectively, and those under the \textit{serial1col} architecture realizing unified encryption/decryption with and without FOMHIDTDUP countermeasures, called \textit{serial1col\_uni} and \textit{serial1col\_unisec}, respectively, were selected to be implemented in the test chip. In these designs, $k$ and $l$ were set to 128. In addition to the Kalyna and AES crypto cores, the test chip comprises a serial interface with control and status registers (CSRs) that facilitate data transfer to and from outside the chip, as well as a controller module that (i) orchestrates the start of operation for a chosen cryptographic block; (ii) performs the writing of results to a serial CSR; and  (iii) signals that the chip is ready for a new operation. The controller module is clocked by an external ``slow'' 10 MHz clock pin (\texttt{clock\_slow\_i}), while the cryptographic blocks can be clocked by either the same slow clock or a faster clock on the ``fast'' clock pin (\texttt{clock\_fast\_i}). The slow clock is used for initial cryptographic block correctness validation, while the fast clock is used both for correctness validation and power measurements. Fig.~\ref{fig:chip_block_diagram} presents the block diagram of the test chip structure.

\begin{figure}[t]
\centering
    \begin{tikzpicture}[spy using outlines={circle,black,dashed,thick, magnification=3.5,size=3.0cm, connect spies}]
        \node {\includegraphics[width=0.80\columnwidth, angle=270, trim=700 0 850 100, clip]{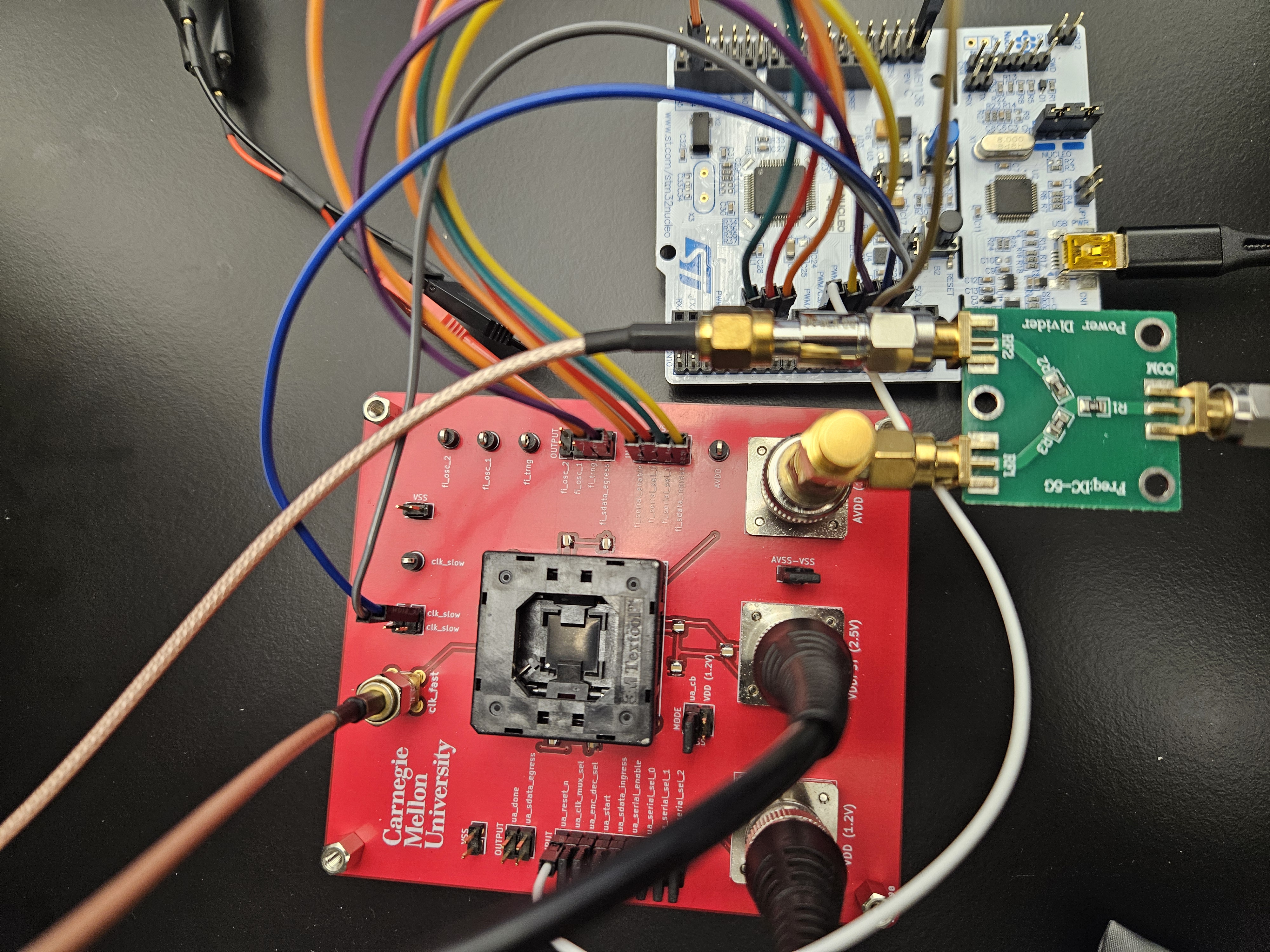}};
        \node[draw, fill=white, font=\small] (node1) at (-0.8,0.9){ASIC};
        \draw[->,red,very thick](node1) -- ++(-0.5,-0.5);
        
        \node[draw, fill=white, font=\small] (node2) at (3.2,0.0){STM32};
        \draw[->,red,very thick](node2) -- ++(-0.5,-0.5);

        \node[draw, fill=white, font=\small] (node3) at (-1.25,2.6){Clock delivery};
        \draw[->,red,very thick](node3) -- ++(-0.5,-0.5);

        \node[draw, fill=white, font=\small] (node4) at (-2.8,-2.6){Power delivery};
        \draw[->,red,very thick](node4) -- ++(+0.5,+0.5);
        
    \end{tikzpicture}
    \caption{Test chip measurement setup using an STM32 controller to drive the ASIC chip.}
    \label{fig:lab}
\end{figure}
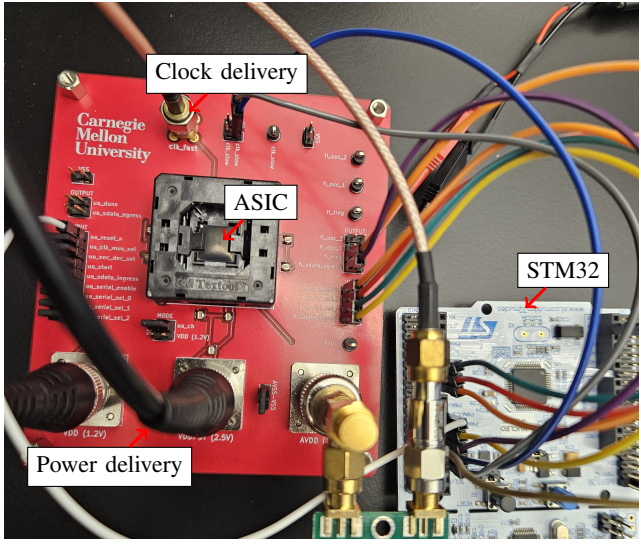

Fig.~\ref{fig:lab} shows the bench measurement setup used in the bring-up process of the test chip. An STM32 board (white) is used to read and write values to and from the serial CSRs, as well as set control bits such as operation start and clock selection. An RF sine wave generator is used to drive the "fast" clock pin.

Table~\ref{tab:freqs} shows the sign-off frequencies at the worst possible corner (\textit{SS\_1.08V\_125C}) and measured frequencies at 1.2V and room temperature. The clock frequencies for the cryptographic blocks in the test chip were chosen in the following manner: AES, \textit{serial1opr\_enc}, and \textit{serial1col\_unisec} blocks were assigned their maximum clock frequency based on synthesis experiments with tentative clock periods being reduced at 0.25\;ns steps until negative setup timing slack was observed. The \textit{serial1col\_uni} and \textit{serial1opr\_uni} blocks were assigned the 308\;MHz maximum sign-off clock frequency to match \textit{serial1col\_unisec}. An average of 1.4x gain in measured maximum clock frequency versus sign-off frequency is observed, due to the corner being considered for setup timing pessimism. 


Comparing the similar AES and \textit{serial1opr\_enc} implementations, the AES design shows a $1.2\times$ higher maximum clock frequency, due to the modulo-64 adder and 4 different S-boxes necessary for Kalyna that are not required in the AES algorithm. The difference between measured and sign-off frequencies for the \textit{serial1col\_uni}, \textit{serial1opr\_uni} and \textit{serial1col\_unisec} implementations can be explained by the complexity between these blocks as demonstrated by the difference in area in Section~\ref{sec:results}, with the less complex blocks showing a higher measured frequency also in silicon.

\begin{table}[t]
    \centering
    \caption{Cryptographic Core Maximum Measured vs Signoff Frequencies (MHz).}
    \vspace*{-3mm}
    \begin{tabular}{|l|ccc|}
        \hline
        Crypto Core & Measured Freq. & Signoff Freq. & Cell Area (um$^2$)\\ 
        \hline \hline
        AES             & 1160 & 800 & 24722.64 \\ 
        serial1opr\_enc & 920 & 666 & 82950.12 \\ 
        \hline
        serial1col\_uni    & 540 & 308 & 43906.32 \\ 
        serial1opr\_uni    & 500 & 308 & 62028.36 \\ 
        serial1col\_unisec & 420 & 308 & 146856.6 \\ 
        \hline
    \end{tabular}
    \vspace*{-4mm}
    \label{tab:freqs}
\end{table}

\begin{figure*}[h]
    \centering
    \includegraphics[width=1\linewidth]{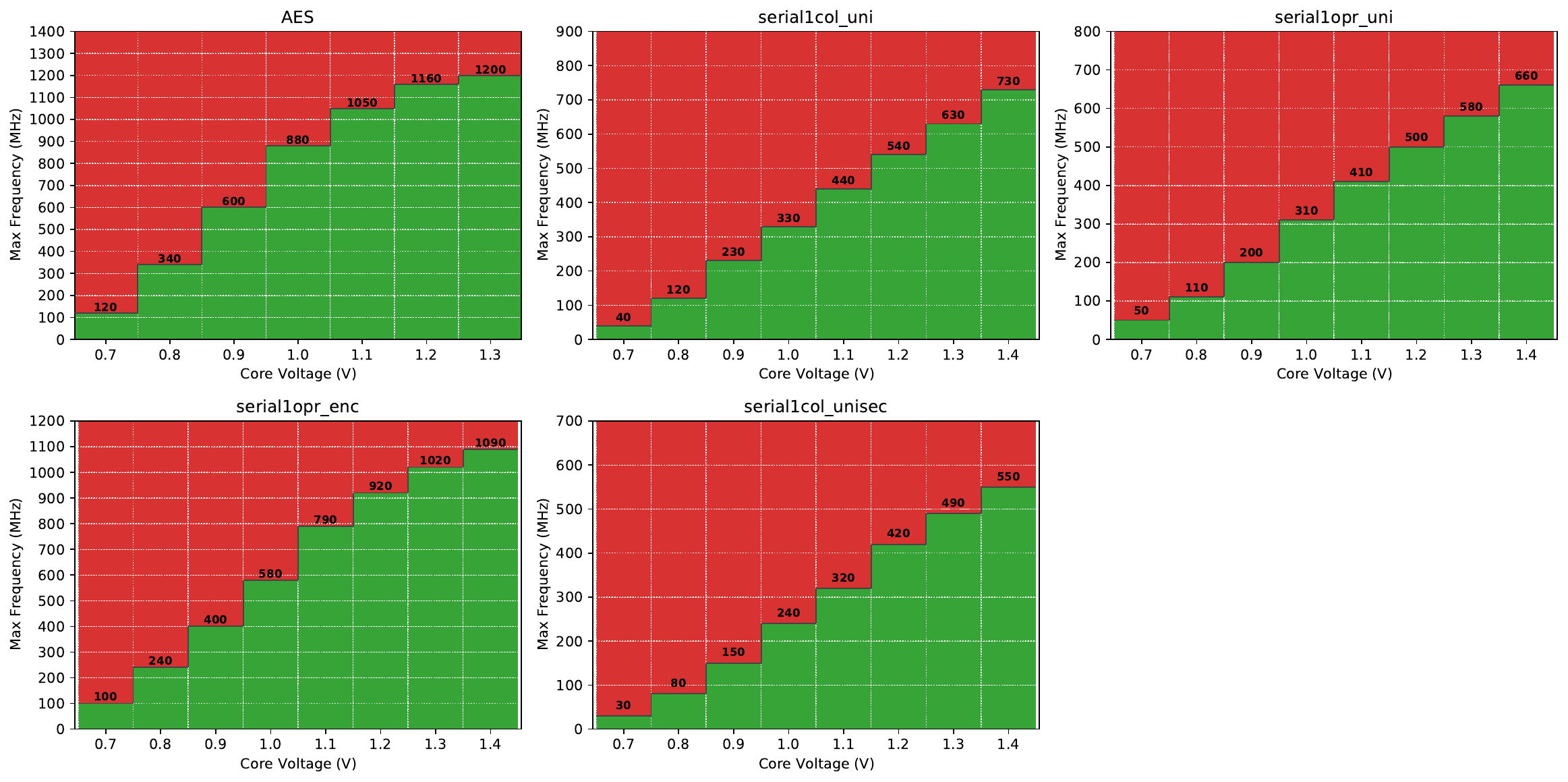}
    \vspace*{-4mm}
    \caption{The shmoo plots of the test chip with various cryptographic blocks.}
    \label{fig:shmoo}
\end{figure*}

Fig.~\ref{fig:shmoo} presents shmoo plots for all Kalyna and AES implementations in the test chip. The maximum frequency for each supply voltage level is determined by raising the tentative clock frequency by 10 MHz until failures are observed. The shmoo curves predictably follow a steady increase in maximum frequency as supply voltage is increased. However, this is not observed on the upper end of the AES shmoo plots, where higher supply voltages do not translate to a linearly increasing maximum frequency. This is attributed to clipping/duty distortion at the IO cell that receives the fast clock signal rather than the behavior intrinsic to the AES crypto core itself.

Table~\ref{tab:energy} shows measured energy costs for an encryption operation in each of the Kalyna implementations, and AES for comparison, as well as the leakage power for each one of the cryptographic cores. 
Measurements were taken considering a 1.2\;V supply voltage and a 200 MHz clock frequency. 
Latency values consider added cycles due to test chip wrappers.
Comparing the encryption-only AES implementation with a 32-bit datapath and on-the-fly key schedule expansion with the closest similar encryption-only Kalyna architecture (\textit{serial1opr\_enc}), the Kalyna encryption operations are found to be $4.12\times$ more costly in terms of energy, and $4.61\times$ more leaky in terms of static power consumption. 
Once again, this can be attributed to the depth of the Kalyna datapath.

\begin{table}[t]
    \centering
    \caption{Energy per block encryption @ 1.2 V / 200 MHz.}
    \vspace*{-3mm}
    \begin{tabular}{|l|ccc|}
        \hline
        Crypto Core & \# Cycles & Energy (nJ) & Leakage (uW) \\ 
        \hline \hline
        AES & 54 & 1.27 & 11.10 \\ 
        serial1opr\_enc & 66 & 5.53 & 51.17 \\ 
        \hline
        serial1col\_uni & 192 & 8.180 & 4.32 \\ 
        serial1opr\_uni & 66 & 6.540 & 8.41 \\ 
        serial1col\_unisec & 381 & 11.57 & 48.25 \\ 
        \hline
    \end{tabular}
    \label{tab:energy}
    \vspace*{-6mm}
\end{table}

Table~\ref{tab:area} shows the total area and instance count for each crypto core, in terms of combinational, sequential, integrated clock gating (ICG), buffer and inverter cells. Note that the buffer and inverted logic overlap with the combinational logic. Contrasting the AES and \textit{serial1opr\_enc} crypto cores, the more complex nature of the Kalyna algorithm lends itself to a more costly implementation than AES. This disparity in algorithm complexity amounts to a $3.35\times$ bigger \textit{serial1opr\_enc} implementation, in addition to the previously mentioned $1.26\times$ difference in maximum frequency in favor of AES. Comparing the three 308\;MHz implementations, the physical protection features of \textit{serial1col\_unisec} produce a $2.37\times$ bigger implementation when compared to the similar \textit{serial1col\_uni} design. This difference is even greater in comparison to the architecturally dissimilar \textit{serial1opr\_uni} implementation, which translates to a $1.41\times$ bigger implementation with $2.9\times$ better latency.

\begin{table*}[!h]
    \centering
    \caption{Crypto Core Instance Summary - \textit{Cell Area (\# Instances).}}
    \begin{tabular}{|l|cccccc|}
    \hline
    Crypto Core (Signoff Freq.)    & Total            & Sequential            & Combinational     & ICG         & Buffer         & Inverter         \\ 
    \hline \hline
    AES (800 MHz)            & 24722.64 (8201)  & 2840.4 (288)    & 21882.24 (7913)   & 72.72 (7)   & 1294.56 (698)  & 1979.28 (1403)   \\
    serial1opr\_enc (666 MHz) & 82950.12 (25046) & 6725.52 (788)   & 76224.6 (24258)   & 59.76 (7)   & 3263.76 (1690) & 8716.68 (4948)   \\ 
    \hline
    serial1opr\_uni  (308 MHz)    & 62028.36 (23996) & 6539.04 (788)   & 55489.32 (23208)  & 54.72 (7)   & 1437.84 (638)  & 5268.24 (4446)   \\
    serial1col\_uni  (308 MHz)    & 43906.32 (15346) & 11641.32 (1448) & 32265.0 (13898)   & 118.44 (16) & 1084.32 (471)  & 2485.44 (2132)   \\
    serial1col\_unisec (308 MHz) & 146856.6 (56542) & 14745.96 (1787) & 132110.64 (54755) & 160.92 (21) & 6138.72 (2780) & 13482.36 (10572) \\
    \hline
    \end{tabular}
    \label{tab:area}
\end{table*}

	\section{Conclusions}
\label{sec:conclusions}

This paper presents the design space exploration of the Kalyna block cipher targeting an ASIC design, introduces alternative design architectures to find the trade-off between area and latency, and proposes hardware reduction techniques. It explores Kalyna implementations with all possible block size and key length combinations, realizing all possible functions, i.e., encryption, decryption, and unified encryption/decryption. It also introduces hardware-efficient implementations of the Kalyna block cipher, including countermeasures against SCA and FI attacks, such as hiding, first-order masking, and duplication, and presents the power leakage of these designs. Furthermore, it presents the results from the ASIC chip realizing various Kalyna designs.

Experimental results show that the proposed architectures lead to alternative designs with area, latency, and energy consumption that can satisfy the design requirements in a specific application. The security countermeasures significantly increase the hardware complexity for the sake of resiliency against SCA and FI attacks. The Kalyna designs in the developed ASIC test chip behave as expected, validated through simulation.


    \section*{Acknowledgments}

    This work was supported by the National Science Foundation under Award No. 2414083, by the U.S. National Academy of Sciences (NAS) and the Office of Naval Research Global (ONRG) through the Science and Technology Center in Ukraine (STCU) under STCU project number 7127 in the framework of the International Multilateral Partnerships for Resilient Education and Science System in Ukraine (IMPRESS-U), and by the Estonian Research Council through the funding of the international multilateral partnership IMPRESS-U project “EAGER: Hardware-Efficient Realization of UA Cryptographic Standards”.
	\bibliography{kalyna}
	\bibliographystyle{IEEEtran}
	
\end{document}